\documentclass[preprint,12pt]{elsarticle}

\usepackage{amssymb}
\usepackage{amsmath} 
\usepackage{mathtools} 
\usepackage{IEEEtrantools} 
\usepackage{mathrsfs} 
\usepackage{float}
\usepackage[margin=2.0cm]{geometry}

\journal{Acta Materialia}
\begin{document}

\begin{frontmatter}

\title{Statistical analysis of dislocation cells in uniaxially deformed copper single crystals}

\address[1]{Eötvös Loránd University, Department of Materials Physics, 1117 Budapest, Pázmány Péter sétány 1/a, Hungary}

\address[2]{Empa, Swiss Federal Laboratories for Materials Science and Technology, Laboratory of Mechanics of Materials and Nanostructures, CH-3602 Thun, Feuerwerkerstrasse 39., Switzerland}

\address[3]{CNRS, UMR 5307 LGF, Mines Saint-Étienne, Centre SMS, 158 cours Fauriel 42023 Saint-Étienne, France}

\address[4]{Thin Film Physics Department, Institute of Technical Physics and Materials Science, Centre for Energy Research, Budapest, Hungary}

\author[1]{Sándor Lipcsei}
\author[2,3]{Szilvia Kalácska}
\author[1]{Péter Dusán Ispánovity}
\author[4]{János L. Lábár}
\author[1]{Zoltán Dankházi}

\author[1]{István Groma\corref{cor1}}

\cortext[cor1]{Corresponding author  groma@metal.elte.hu}

\begin{abstract}
The dislocation microstructure developing during plastic deformation strongly influences the stress-strain properties of crystalline materials. The novel method of high resolution electron backscatter diffraction (HR-EBSD) offers a new perspective to study  dislocation patterning. In this work copper single crystals deformed in uniaxial compression were investigated by HR-EBSD, X-ray line profile analysis, and transmission electron microscopy (TEM). With these methods the maps of the internal stress, the Nye tensor, and the geometrically necessary dislocation (GND) density were determined at different load levels. In agreement with the composite model long-range internal stress was directly observed in the cell interiors. Moreover, it is found from the fractal analysis of the GND maps that the fractal dimension of the cell structure is decreasing with increasing average spatial dislocation density fluctuation. It is shown that the evolution of different types of dislocations can be successfully monitored with this scanning electron microscopy based technique.

\end{abstract}



\begin{keyword}
Dislocation structures \sep Electron backscatter diffraction (EBSD) \sep X-ray diffraction \sep Dislocation patterning
\end{keyword}

\end{frontmatter}

\section{Introduction}
It was first observed nearly 60 years ago that dislocations created during the plastic deformation of crystalline materials tend to form different patterns with morphology depending on the mode, temperature and rate of deformation. There is an equally longstanding discussion regarding the physical origin of these patterns. A large variety of approaches have been proposed to model the instability leading to the spatial variation of the dislocation density, many of which are based upon analogies with pattern formation in other physical systems. It has been argued that dislocation patterns can be understood by the tendency toward the minimization of some kind of elastic energy functional (Hansen and Kuhlmann-Wilsdorf \cite{hansen1986low}, Holt \cite{holt1970dislocation}, Richman and Vinas \cite{rickman1997modelling}), but the theories have never been worked out in details. Another approach proposed is to model the dislocation patterning as a reaction-diffusion phenomena of the mobile and inmobile dislocation densities (Walgraef and Aifantis \cite{walgraef1985dislocation}, Pontes et al.~\cite{pontes2006dislocation}). The fundamental problem with this approach is that it is completely phenomenological, i.e.~one can not see how the different terms appearing in the evolution equations are related to the properties of individual dislocations.

In a recent series of papers \cite{groma2016dislocation, wu2018instability, wu2021cell} a new theoretical approach based on a continuum theory of dislocations, derived from the evolution of individual dislocations, was proposed for modelling the patterning process.  According to the theory the main source of the instability is the nontrivial mobility of the dislocations caused by the finite flow stress, while the characteristic length scale of the pattern is selected by the ``diffusion'' like terms appearing in the theory due to dislocation correlation effects. Since, however, the theory is developed for a rather idealised 2D dislocation configuration, further experimental and theoretical investigations are needed to create a general comprehensive theory of dislocation patterning.   

One of the most challenging issue is the characterisation and modelling of the self-similar fractal-like dislocation cell structure formed in FCC crystals oriented for ideal multiple slip (for details see the pioneering works of Zaiser and H\"ahner \cite{hahner1998fractal,zaiser1999flow}). In this paper we present high resolution electron backscatter diffraction (HR-EBSD), X-ray line profile analysis, and transmission electron microscopy (TEM) investigations on compressed Cu single crystals. Since some of the aspects of the applied methods are developed exclusively for the specific requirements of the addressed problem, in the first half of the paper the applied experimental methods are explained in detail. In the second half the obtained results are thoroughly discussed.  

\section{Experimental methods}
\subsection{Sample preparation}
In order to study the dislocation cell formation mechanism in FCC materials a high purity copper single crystal  was used. The initial average dislocation density was confirmed to be $\mathrm{1.5 \times 10^{14}}$ m$^{-2}$ by X-ray line profile analysis (the details of the method applied are explained below in Sec.~\ref{sec:x_ray}). For the compression tests prism shaped samples with dimensions of 2.5 $\times$ 2.5 $\times$ 5 mm$^3$ were cut with an electrical discharge machine (EDM). The orientation of each surface was of (100) type. For removing the amorphous layer created by EDM the specimens were etched in a 30\% $\mathrm{HNO_3}$ solution for 10 minutes. To reduce the initial dislocation density the samples were heat treated at $600 \; \mathrm{^oC}$ for 6 hours in a vacuum furnace. According to X-ray line profile measurements the initial dislocation density decreased to about $\mathrm{2 \times 10^{13} \; m^{-2}}$.

\begin{figure}[H]
  \centering
  \includegraphics[width=0.2475\textwidth]{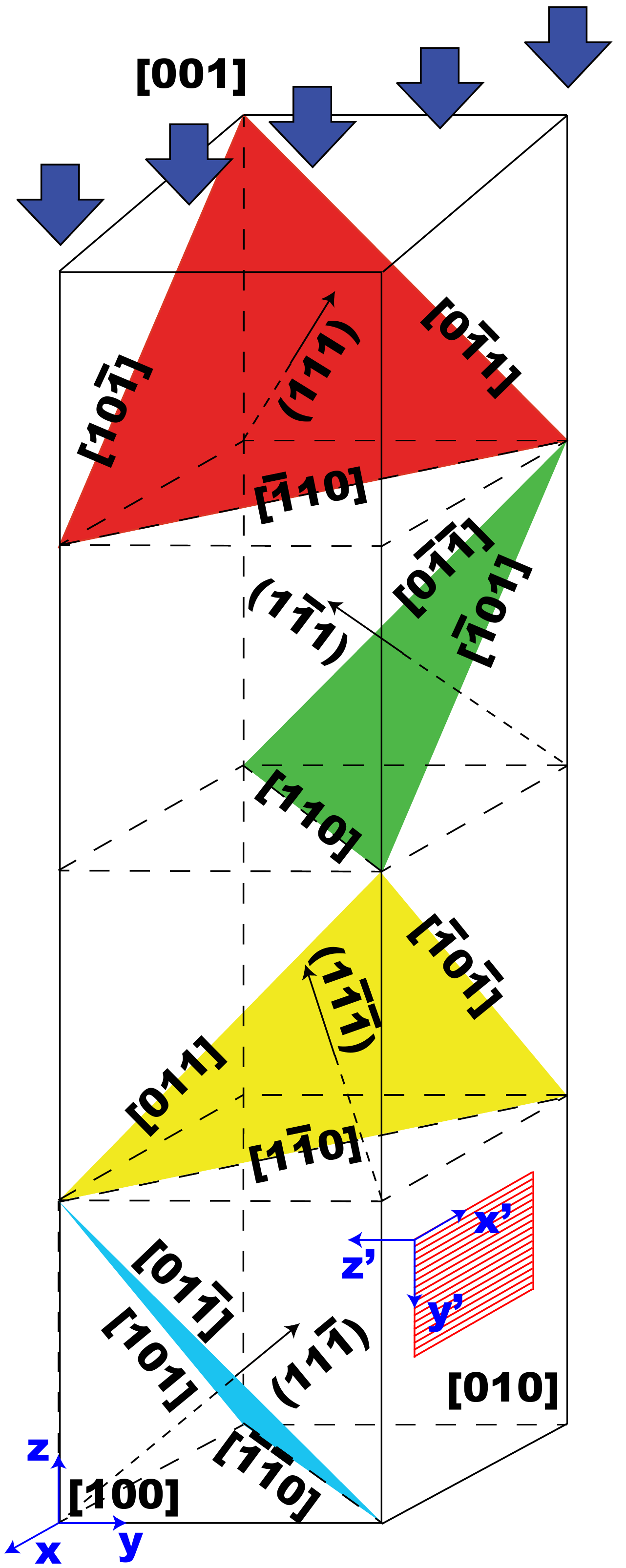}
  \caption{The sample geometry and the active slip planes of the compressed Cu single crystals. The compression was applied on the [001] surface, while the EBSD and X-ray line profile measurements were performed on the [010] surface.}
 \label{geometry}
\end{figure}

The samples were compressed on the squared shape surfaces ensuring  uniaxial deformation in the [001] direction corresponding to ideal multiple slip. The geometry of the compression, EBSD scanning and the activated slip systems are shown in Fig.~\ref{geometry}. 

\begin{figure}[H]
	\centering
		\includegraphics[width=0.7\textwidth]{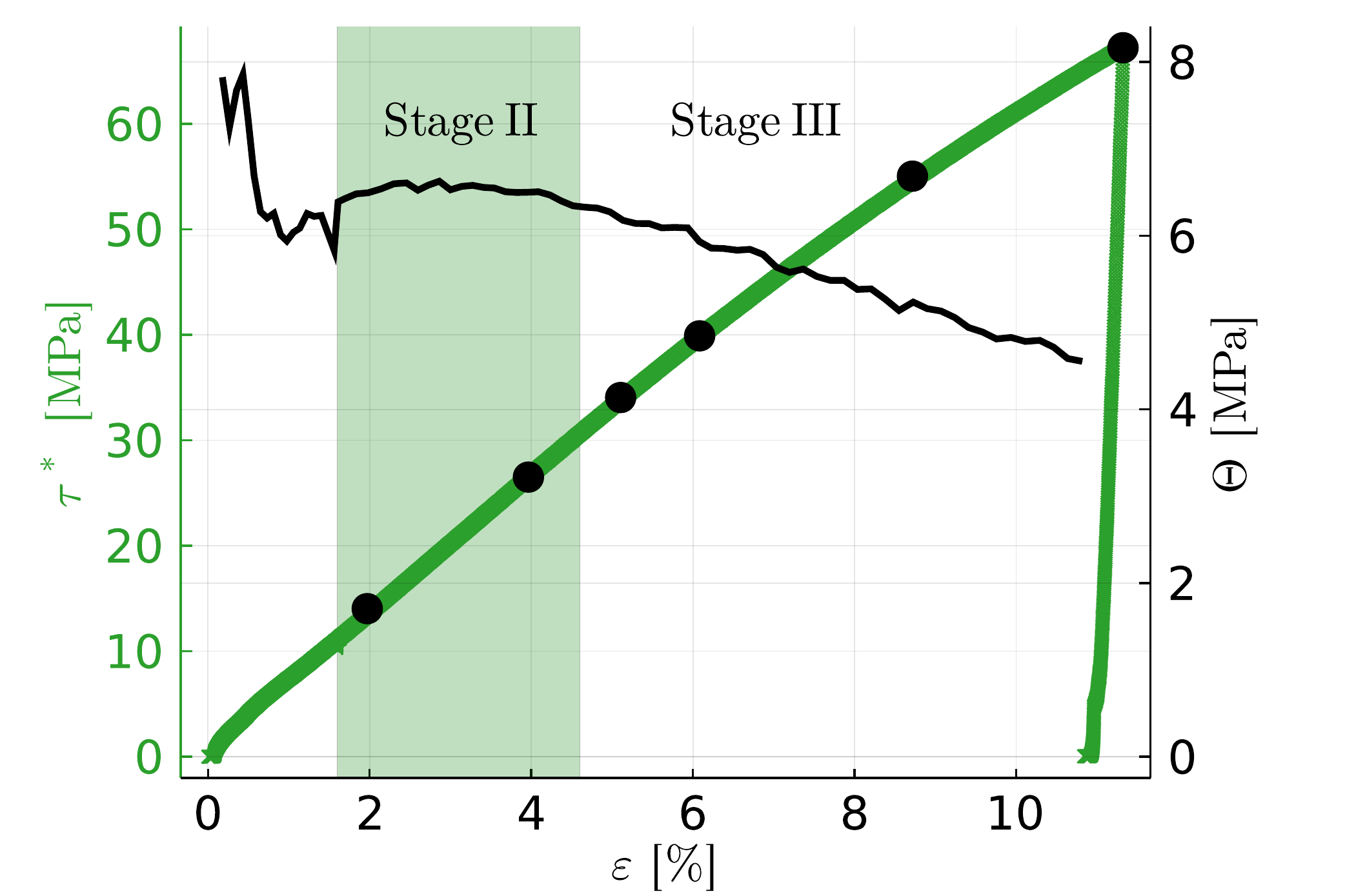}
	\caption{Resolved shear stress and hardening rate vs.~strain obtained on a compressed Cu single crystal oriented for (001) ideal multiple slip. The black dots on the curve mark the stress levels until which the 6 different samples were compressed.}
	\label{StressStrainCurve}
\end{figure}

Six samples were deformed up to different strain levels. The resolved shear stress $\tau^*$ vs.~strain $\varepsilon$ curve of the sample with the highest terminal deformation is shown in Fig.~\ref{StressStrainCurve}. The black dots on the curve mark the maximum stresses and strain levels of the 6 different samples. The red line in the figure shows the hardening rate $\varTheta=d \tau /d \varepsilon$ as a function of strain. As it is expected for ideal multiple slip \cite{workhardening} the $\varTheta(\varepsilon)$ curve consists of a nearly horizontal (stage II) and a decreasing (stage III) part.  

From the six specimens prepared three are close to the strain level corresponding to the transition from stage II to stage III. As it is seen below this strain region is critical for the statistical properties of the dislocation cell structure developing during the deformation.  

Finally, in order to prepare the samples for TEM and HR-EBSD measurements, electropolishing was applied at 20 V, 1.2 A using Struers D2 electrolyte for 30 seconds.

\subsection{X-ray line profile analysis}
\label{sec:x_ray}
\indent X-ray line profile analysis is a well-established method to determine the average dislocation density, the average squared dislocation density and the dislocation polarization from the measured intensity profile. In our analysis the ``restricted moments'' method developed by I.~Groma et al.~\cite{PhysRevB.57.7535,groma1988asymmetric,Ungar:gk0172} was applied. In the evaluation of the measured data the asymptotic behavior of the different order restricted moments are analyzed. The $k\mathrm{^{th}}$ order restricted moments are defined as 
\begin{equation}
    v_k(q) = \frac{\int_{-q}^{q} q'^k I(q') dq'}{\int_{-\infty}^{\infty}I(q')dq'},
\end{equation}
where $I(q')$ is the intensity distribution near to a Bragg peak, in which $q' = 2 (\sin \theta - \sin \theta_0) / \lambda$, $\lambda$ is  the wavelength of the applied X-rays, and $\theta$ and $\theta_0$ are the half of the diffraction and Bragg angles, respectively. 

As it is explained in detail in \cite{PhysRevB.57.7535} 
for large enough $q'$ value the asymptotic form of the $\mathrm{2^{nd}}$ order restricted moment reads as
\begin{equation}
    v_2(q) = 2 \Lambda \langle \rho \rangle \ln{\left(\frac{q}{q_0}\right)},
\end{equation}
were $\langle \rho \rangle$ is the average dislocation density, $q_0$ is a parameter determined  by the dislocation-dislocation correlation, and $\Lambda$ is a constant depending on the dislocation Burgers vector $\vec{b}$, the line direction $\vec{l}$, and the diffraction vector $\vec{g}$. $\Lambda$ is commonly written in the form  $\Lambda=\pi|\vec{g}|^2|\vec{b}^2|C/2$ where $C$ is called the contrast factor. (For its actual value  a detailed deduction and explanation can be found in \cite{PhysRevB.57.7535}.) From the intensity profiles measured the  values of $\Lambda \langle \rho \rangle$ and $q_0$ can be obtained by fitting a straight line on the asymptotic part of the $v_2(q)$ versus $\ln(q)$ plot. 

Beside the $\mathrm{2^{nd}}$ order restricted moment for our analysis the $\mathrm{4^{th}}$ order restricted moment is also important. In the asymptotic regime it is \cite{PhysRevB.57.7535}:
\begin{equation}
    v_4(q) = \Lambda \langle \rho \rangle q^2 + 12 \Lambda^2 \langle \rho^2 \rangle \ln^2{\left(\frac{q}{q_1}\right)},
    \label{v4}
\end{equation}
where $\langle \rho^2 \rangle $ is the average dislocation density fluctuation, and $q_1$ is a parameter. For the better visualization it is useful to consider the quantity
\begin{equation}
    \frac{v_4(q)}{q^{2}} = \Lambda \langle \rho \rangle + 12 \Lambda^2 \langle \rho^2 \rangle \frac{\ln^2{\left(\frac{q}{q_1}\right)}}{q^{2}},
    \label{eqf}
\end{equation}
which asymptotically tends to $\Lambda \langle \rho \rangle$. The actual values of the parameters $\Lambda \langle \rho \rangle$, $\Lambda \langle \rho^2 \rangle$, and $q_1$ can be determined by fitting the form given by Eq.~(\ref{eqf}) to the asymptotic regime of the $ v_4(q)/q^2 $ versus $q$ plot.

An important statistical parameter of the dislocation microstructure developed is the relative dislocation fluctuation defined as
\begin{equation}
    \sigma = \sqrt{\frac{\langle \rho^2 \rangle - \langle \rho \rangle^2}{\langle \rho \rangle^2}} 
    \label{avg.dis.dens.fluct.}
\end{equation}
that can be determined from the $\mathrm{4^{th}}$ order restricted moment. 

It should be noted that the measured intensity often contains a background which has to be subtracted  before the calculation of the restricted moments. Since however, the background has different contribution to the $\mathrm{2^{nd}}$ and $\mathrm{4^{th}}$ order restricted moments, determining the average dislocation density from both moments offers a internal checking possibility whether the background level was selected correctly. 

The profile measurements have been performed with a Cu rotating anode Cu X-ray generator at 40 kV and 100 mA  with wavelength $\lambda =0.15406$ nm. In order to reduce the instrumental broadening the symmetrical (220)
reflection of a Ge monochromator was used. The K$\alpha_2$ component of the Cu radiation was eliminated by an 0.1 mm slit between the source and the Ge crystal. The profiles were registered by a linear position sensitive DECTRIS MYTHEN2 R detector with $50$ $\mu$m spatial resolution and 1280 channels. The sample-detector distance was $0.7$ m resulting an  angular resolution  in the order of  $0.004^{\circ}$.

\begin{figure}[H]
	\centering
		\includegraphics[width=0.7\textwidth]{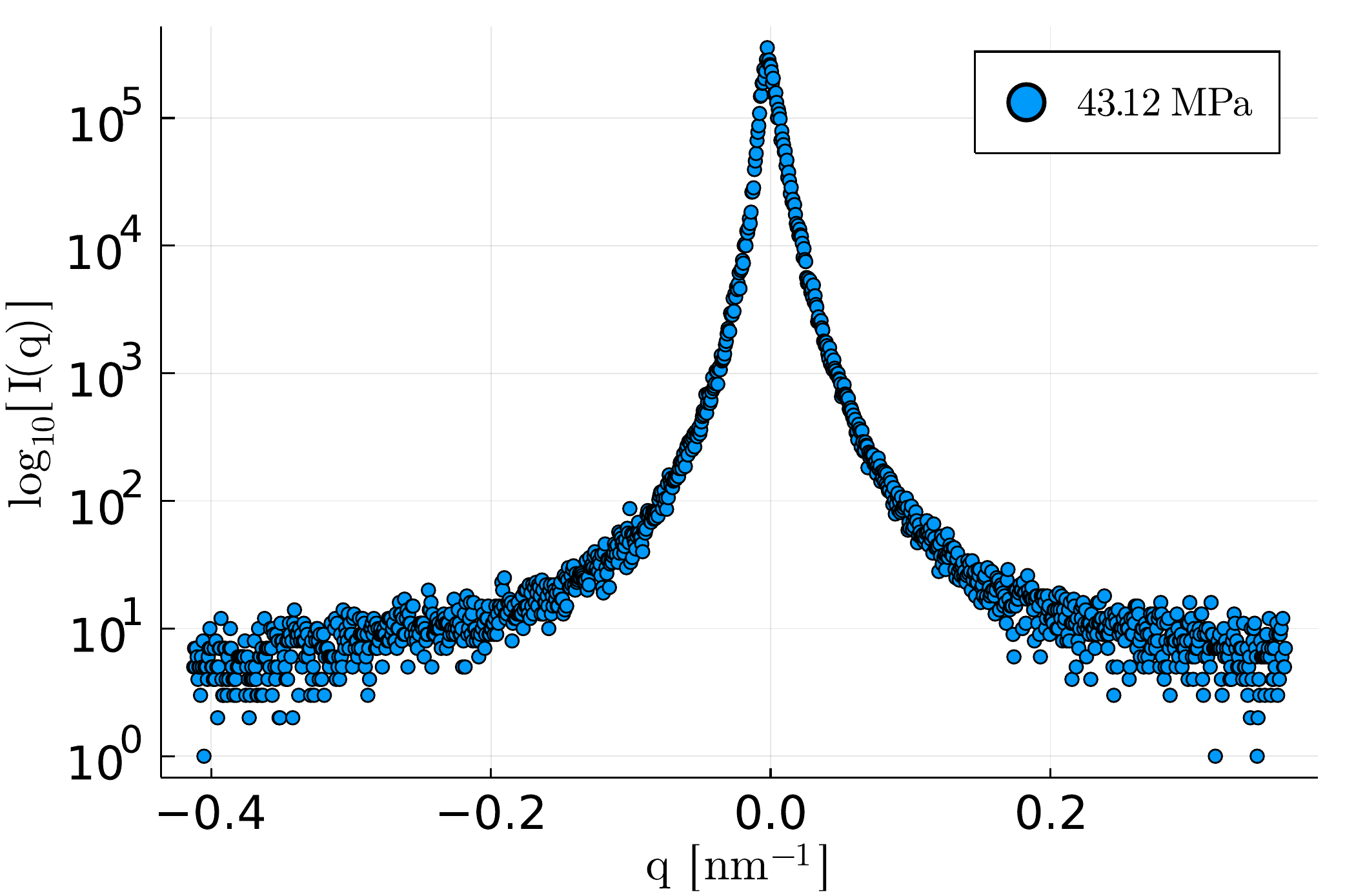}
	\caption{The X-ray line profile obtained at $\vec{g}=(020)$ on the sample compressed up to 43.12 MPa. In order to eliminate the effect of the noise the peak intensity should be at least $\mathrm{10^3-10^4}$ times higher than the background, and a subsequent background subtraction should be carried out.}
	\label{IntDist}
\end{figure}

\begin{figure}[H]
	\centering
		\includegraphics[width=0.7\textwidth]{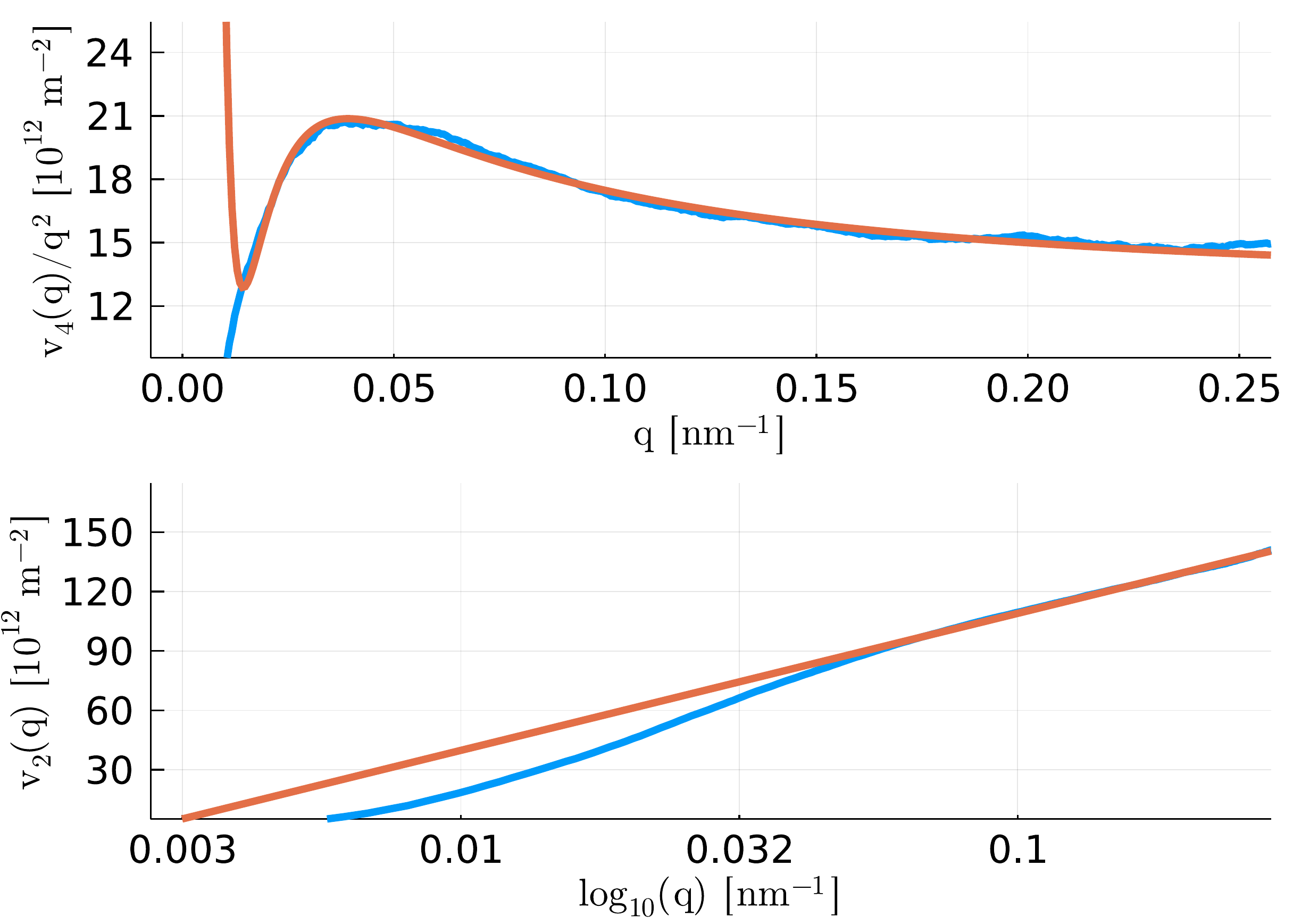}
	\caption{The raw data (blue lines) and the fitted restricted moments (orange lines).}
	\label{ExampleM2M4}
\end{figure}

The evaluation method applied is demonstrated on the intensity distribution (Fig.~\ref{IntDist}) obtained on the sample compressed up to 43.12 MPa resolved shear stress and on the corresponding restricted moments (Fig.~\ref{ExampleM2M4}). The different parameters can be determined with an accuracy of less than 5\%.

\subsection{HR-EBSD}

EBSD measurements were carried out in a FEI Quanta 3D SEM equipped with an Edax Hikari EBSD detector. Diffraction patterns were recorded with 1$\times$1 binning (640 px $\times$ 480 px resolution) using an electron beam of 20 kV, 16 nA. In order to carry out statistical analysis on the collected data a 20 $\mu$m $\times$ 20 $\mu$m area was mapped with a step size of 100 nm on each sample. The HR-EBSD technique utilizes image cross-correlation on the recorded diffraction patterns \cite{Wilkinson.2012}. The local strain tensor components can be determined, and a lower bound estimate of the GND density can be given using the commercially available software. The method requires an ideally stress-free diffraction pattern as reference, that is often difficult to obtain experimentally. In the absence of such reference, it is noted that the scales should be implemented as relative and not absolute measures. Image cross-correlation based HR-EBSD calculations were performed using BLG Vantage CrossCourt v.4 software that provided the components of the elastic distortion ($\beta_{ij}^{el}$) and the stress tensor ($\sigma_{ij}$) and the also the values of the GND density ($\rho_\mathrm{GND}$).

From the distortion map the Nye dislocation density tensor $\alpha_{ij}$ defined as \cite{kroner1981continuum}
\begin{eqnarray}
 \alpha_{ij}=-e_{klj}\partial_k \beta^{el}_{il}
\end{eqnarray}
can be determined where $\beta_{ij}^{el}$ is the elastic distortion tensor and $e_{ijk}$ is the Levi-Civita symbol. Since, however, in the HR-EBSD measurement the distortion tensor is measured directly on the sample surface, only those components of $\alpha_{ij}$ can be calculated that are independent from the derivation in the direction perpendicular to the sample surface. So, in a coordinate system with $z$ axis perpendicular to the sample surface only the ${iz}$ components of the Nye tensor
\begin{eqnarray}
 \alpha_{iz}=\partial_y \beta^{el}_{ix}-\partial_x \beta^{el}_{iy}
\end{eqnarray}
can be directly determined from a HR-EBSD measurement.
Since
\begin{eqnarray}
 \alpha_{ij}=\sum_t b_i^t l_j^t \rho^t \label{eq_alpha}
\end{eqnarray}
where the superscript $t$ denotes a given type of dislocation present in the system with Burgers vector $\vec{b}^t$, line direction $\vec{l}^t$, and dislocation density $\rho^t$, from the measured Nye tensor components, one can make an estimate on the dislocation population in the different slip systems (for details see below). 

Furthermore, to characterize the GND density the scalar quantity
\begin{eqnarray}
 \rho_\mathrm{GND}=\frac{1}{b}\sqrt{\alpha_{xz}^2+\alpha_{yz}^2+\alpha_{zz}^2} \label{eqGND}
\end{eqnarray}
was introduced. The GND density and the $\alpha_{iz}$ tensor components were determined using a C++ code developed by some of the Authors \cite{Kalacska.2017, Kalacska.2020a}.

\section{Stress-map analysis with the momentum method}

\begin{figure}[H]
  \centering
  \includegraphics[width=0.325\textwidth]{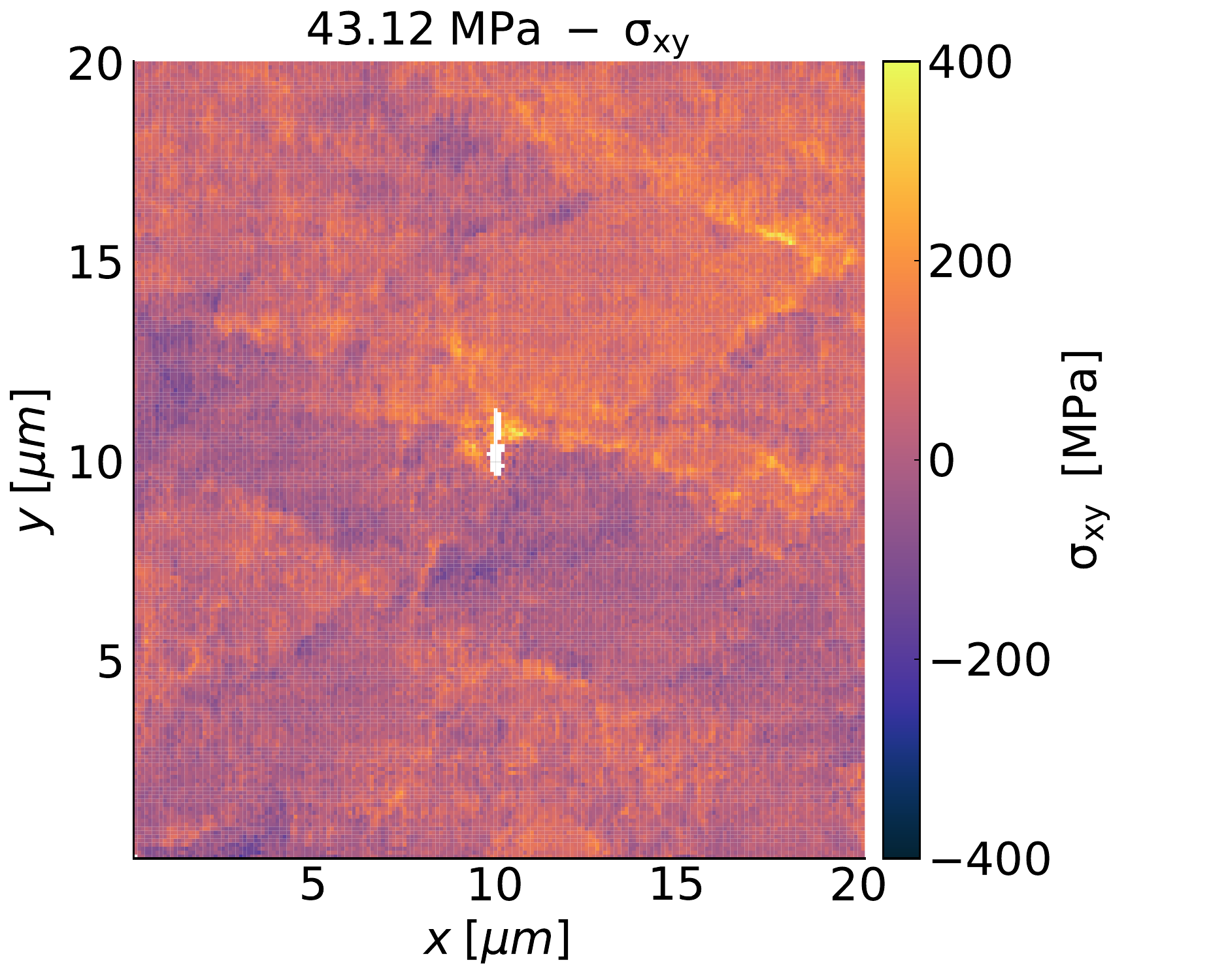}
  \includegraphics[width=0.325\textwidth]{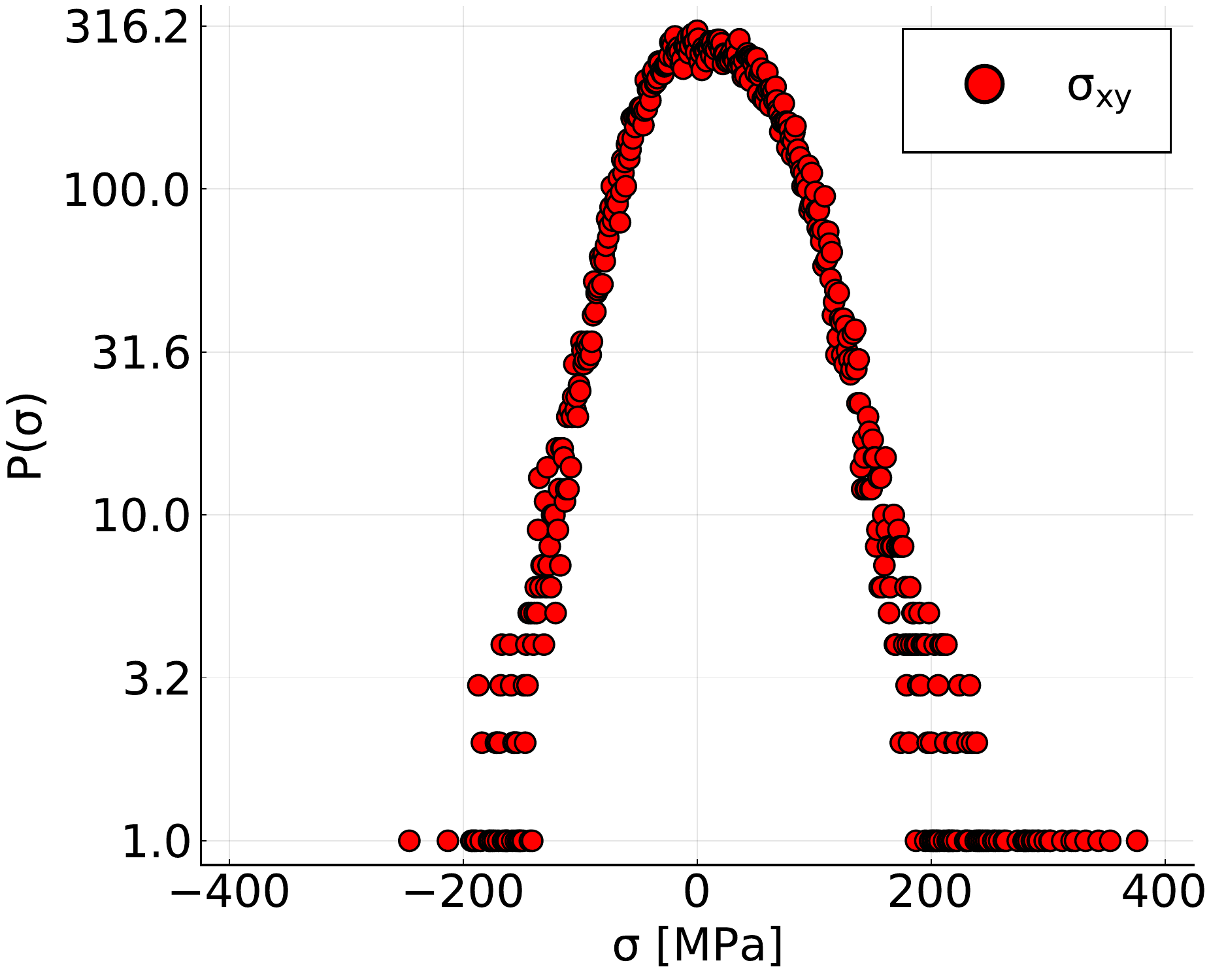}
  \includegraphics[width=0.325\textwidth]{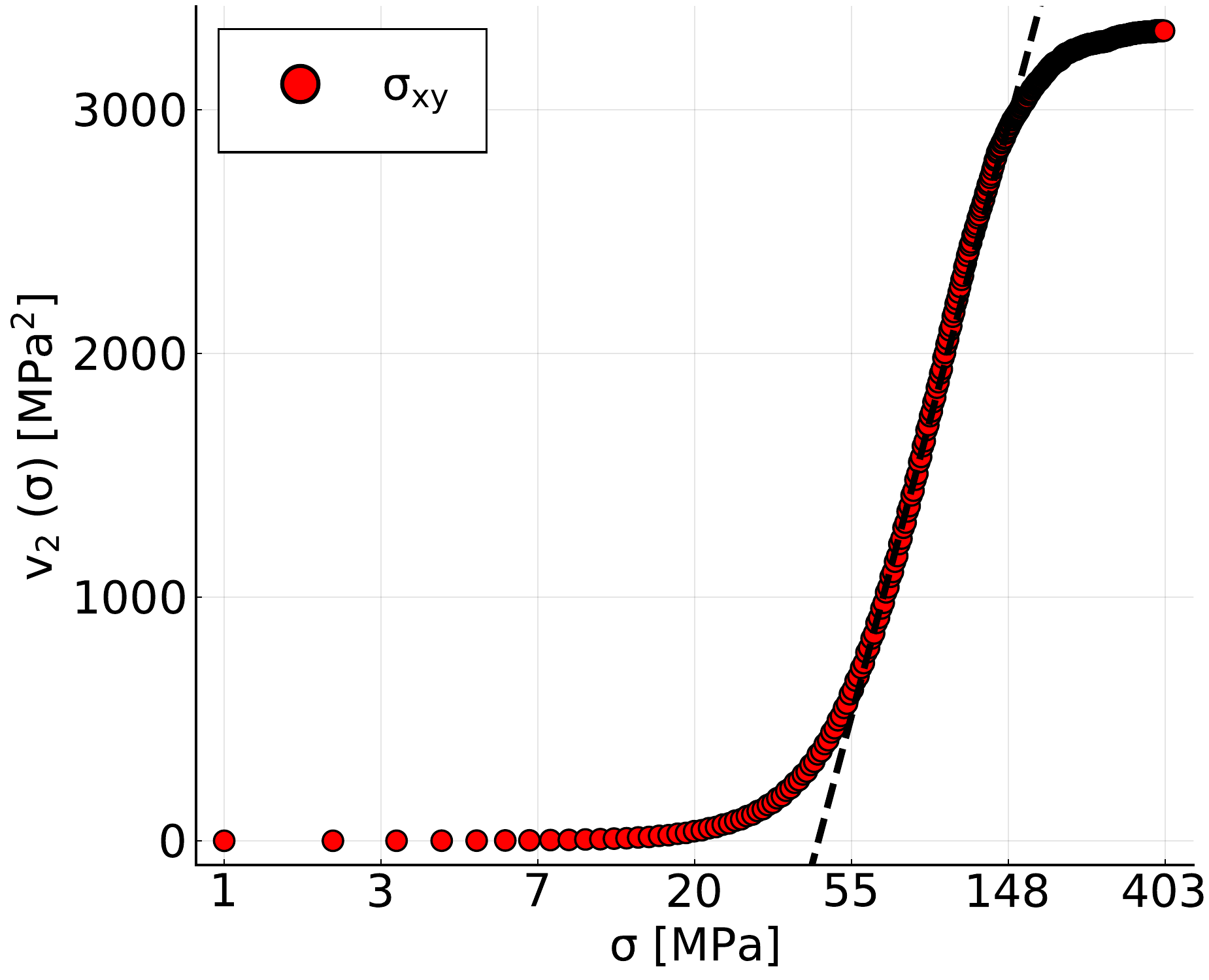}
	\caption{Stress map, stress probability distribution, and the corresponding second order restricted moment obtained on the sample compressed up to $\mathrm{43.12 \; MPa}$. }
	\label{stress_plot}
\end{figure}

It was already demonstrated earlier by Groma et al. and Wilkinson et al. \cite{groma1998probability, csikor2004probability, wallis2021dislocation}, that for a dislocation ensemble of parallel edge dislocations the asymptotic part of the probability distribution of the internal stress $p(\sigma)$ decays as   
\begin{equation}
    p(\sigma) \approx  \frac{b^2 \mu^2}{8 \pi^2} C \langle \rho \rangle \frac{1}{\sigma^3}
\end{equation}
 were  $\mu$ is the shear modulus and $C$ is a ``geometrical'' constant depending on the type of dislocation similar to the contrast factor in the case of X-ray peaks, and the stress component considered. So, like for X-ray line broadening, in the asymptotic regime the second order restricted moment of $p(\sigma)$ is linear in $\ln(\sigma)$. It should be noted, however, that the stress value obtained by HR-EBSD in a given scanning point is the average stress on the area illuminated by the incoming electron beam. As a result at large enough stress levels the probability distribution $P(\sigma)$ measured deviates from the inverse cubic decay, as it turns to a much faster decaying regime \cite{Kalacska.2017}. Nevertheless, for most cases one can easily identify a linear regime on the the second order restricted moment of $p(\sigma)$ versus  $\ln(\sigma)$ plot (see Fig.~\ref{stress_plot}).
 From the deviation of the inverse cubic decay we can define a characteristic length scale $r_d=\mu b /\sigma_d$ where $\sigma_d$ is the stress level where the probability distribution start to deviate from the inverse cubic regime. In the investigations performed $r_d\approx75$ nm. This means, that dislocation dipoles narrower than $r_d$ are not ``seen'' by this method. So, compared to X-ray line profile analysis HR-EBSD somewhat underestimates the dislocation density (for details see below).  
 
 Since from the HR-EBSD analysis one can obtain 5 independent stress components ($\sigma_{zz} \equiv  \sigma_{33}$ is assumed to vanish in the HR-EBSD analysis) a ``formal'' dislocation density $\rho^*_{ij}=b^2 \mu^2/ (8\pi) C_{ij} \langle \rho \rangle$ can be determined from the stress maps corresponding to different $ij$ stress components, where the parameter $C_{ij}$ is the geometrical constant of the $ij^{\mathrm{th}}$ stress component. Unlike for the X-ray line broadening there is no existing analytical calculation to give the precise value for $C_{ij}$. ($C_{ij}$ is calculated only for the shear stress generated by edge dislocations in isotropic materials in the coordinate system defined by the Burgers and line direction vectors of the dislocation \cite{groma1998probability}. For the shear stress $C_{12}=\pi/[2(1-\nu)^2)]$ where $\nu$ is the Poisson's ratio.) In the results presented we gave only the average of the 5 formal dislocation densities.      
 
 A typical stress map, stress probability distribution, and the corresponding second order restricted moment can be seen in Fig.~\ref {stress_plot}.

\subsection{TEM investigations}

A TEM specimen was fabricated from the bulk copper single crystal deformed up to $\mathrm{43.12 \; MPa}$ resolved shear stress with the aim of qualitative comparison of dislocation structures with those obtained from GND density maps. The TEM lamella preparation was carried out via a FEI Quanta 3D FEG dual-beam SEM-FIB microscope. The initial fabrication process was carried out at 30 kV acceleration voltage and 1,3,5 and 30 nA ion current. The finishing consisted of low current (220 pA, 470 pA) and low voltage (2 kV, 5 kV) ion polishing. It is noted that to be able to investigate very large dislocation cells unusually large (20 $\mu$m $\times$ 20 $\mu$m) specimens were fabricated requiring extra care during the preparation process \cite{mayer2007tem, giannuzzi2004introduction, yao2007focused}. Bright field images of the dislocation network were recorded on a $6 \times 6$ $\mathrm{cm^2}$ $\mathrm{4k\times4k}$ CETA 16 CMOS camera with $\mathrm{14 \; \mu m}$ pixel size, controlled by VELOX software in a Titan Themis G2 200 transmission electron microscope operated at 200 kV (see  Fig.13).

\section{Fractal analysis}
The dislocation cell structure developing under unidirectional deformation at ideal multiple slip is known to be a ``hole'' fractal \cite{hahner1998fractal}. Since the GND maps obtained by  HR-EBSD measurements allow us to study the dislocation microstructure on a much larger area than one can do with TEM (applied traditionally for microstructure characterization) we performed fractal dimension analysis on the GND maps at different stress levels. We have applied two different methods, the ``traditional'' box counting and the correlation dimension analysis.

\subsection{Box-counting algorithm}

A common algorithm to determine the fractal dimension of a set is the well known box-counting algorithm \cite{salat2017multifractal}.  In the method we cover the image with a $L$ sized grid, and then count the number of boxes $N$ covering part of the image.
The fractal dimension $D_B$ is 
\begin{eqnarray}
 D_B=\frac{d\ln(N)}{d\ln(L)},
\end{eqnarray}
that is obtained by fitting a straight line to the $\ln(N)$ versus $\ln(L)$ plot.
It is a numerically cheap, fast and fairly precise method.

\subsection{Correlation dimension}

One can also measure the geometrical randomness of points through the so-called correlation integral, which may be estimated for large enough systems with the correlation sum \cite{grassberger2004measuring}
\begin{equation}
    C(\epsilon) = \frac{1}{N(N-1)}\sum_{i \neq j}^{N} H \left(\epsilon - |r_i - r_j|\right)
\end{equation}
where $\epsilon$ is the threshold distance, $N$ is the number of non-zero points, $H$ is the Heaviside step function, $r_i$ and $r_j$ are the coordinates of the set points. The correlation integral scales with the threshold distance as \cite{grassberger2004measuring}
\begin{equation}
    C(\epsilon) \propto \epsilon^{D_c},
\end{equation}
where $D_c$ is the correlation dimension. One can easily see, that for points on a circle the correlation dimension $D_c = 1$, for points on a sphere $D_c = 2$ and for points evenly distributed in a sphere $D_c = 3$. For the analysis of 2D embedded geometrical structures one may expect that $1\leq D_c \leq 2$.

\subsection{Image filtering}

The GND maps measured can not be analyzed with the method explained above in a straightforward manner. One issue is that the maps are obviously not binary ones so, one has to introduce some threshold value above which we consider the map intensity to be 1 and 0 below. The fractal dimension obtained may depend on the threshold value chosen. Another problem we face is that the GND map contains numerous random points. They may correspond to individual dislocations or narrow dislocation multipoles but certainly they should not be considered during the fractal analysis.      

A simple method for global binarization is the so-called Otsu's method \cite{otsu1979threshold,lee1990comparative}. (It is analogous to Fischer's Discriminant Analysis \cite{fisher1936use} method and equivalent to a globally optimized k-means clustering method \cite{macqueen1967some, kriegel2017black}.)
In the simplest form it returns a binarised  intensity map threshold by maximizing the inter-class variance.
In order to get this threshold value first the probability distribution of the point intensity $p(I)$ is calculated numerically with some appropriate binning level $L$ chosen. After this with a threshold level $t$ the histogram is cut into two subhistograms separated by the threshold, and the quantity
\begin{equation}
  \sigma_{w}^{2}(t)=P_{0}(t)\sigma_{0}^{2}(t)+P_{1}(t)\sigma_{1}^{2}(t)
\end{equation}
is calculated, where $P_{0}$ and $P_{1}$ are the probabilities of the two classes separated by $t$, while $\sigma _{0}^{2}$ and $\sigma_{1}^{2}$ are variances of the two classes.  The threshold for the image binarization is selected by minimizing $\sigma_w(t)$. 

Otsu's method performs exceptionally well when the histogram obtained on the image has a bimodal distribution and the background and foreground values are separated by a deep valley. However, if the image is corrupted with additive noise or the variation of intensities between background and foreground are large compared to the mean difference, the histogram may degrade. 

\begin{figure}[H]
  \centering
  \includegraphics[width=0.4\textwidth]{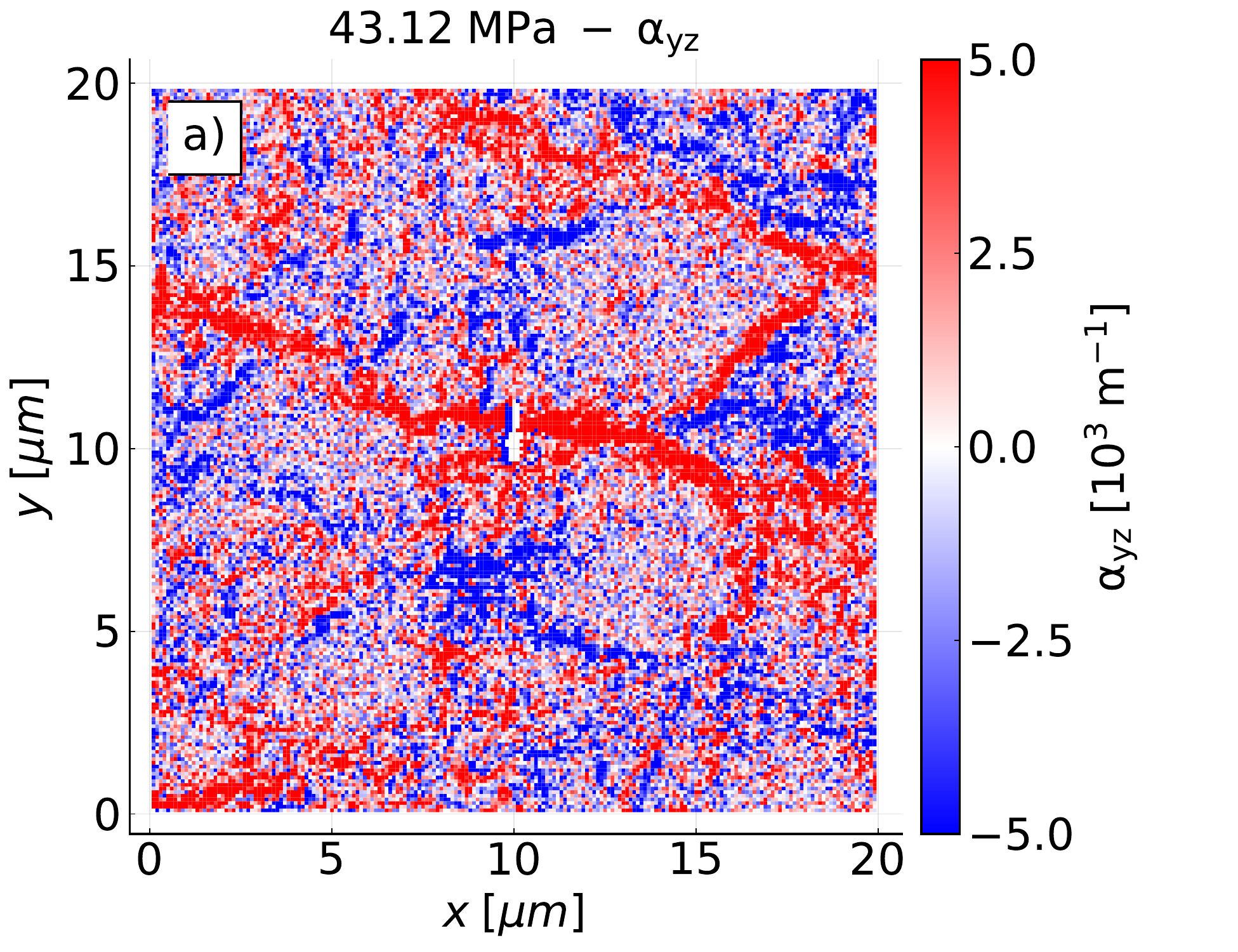}
  \includegraphics[width=0.4\textwidth]{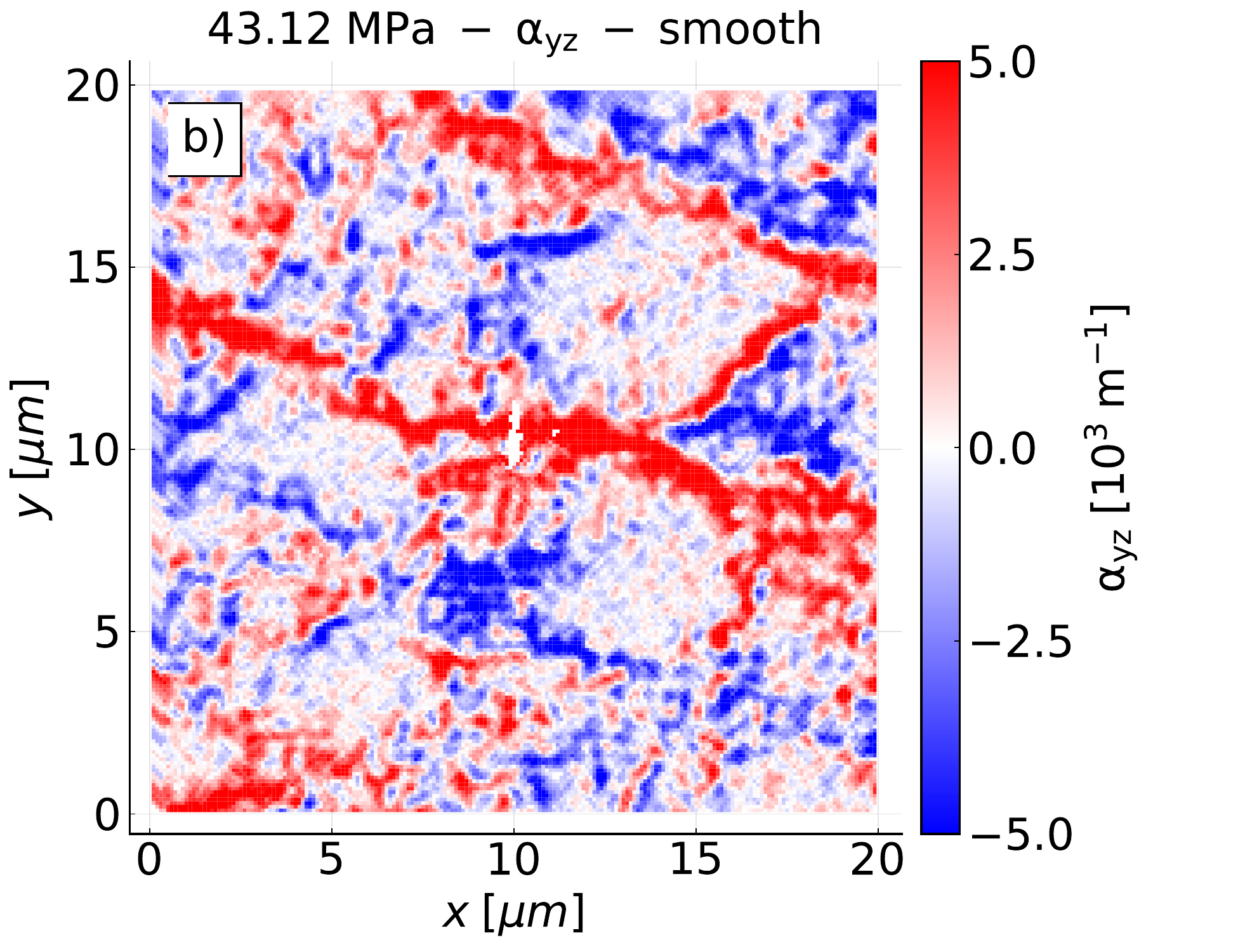}  \\
  \includegraphics[width=0.4\textwidth]{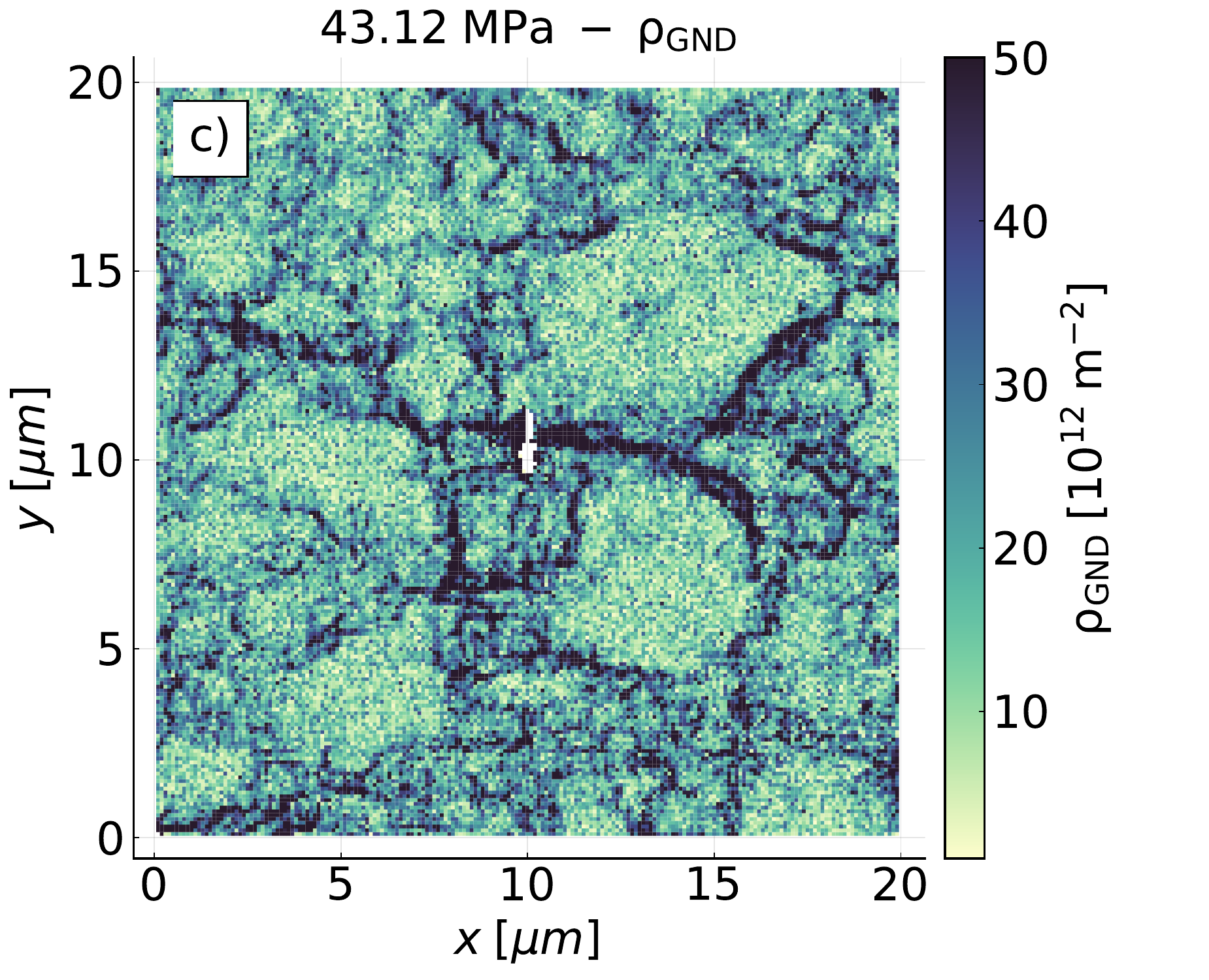}
  \includegraphics[width=0.4\textwidth]{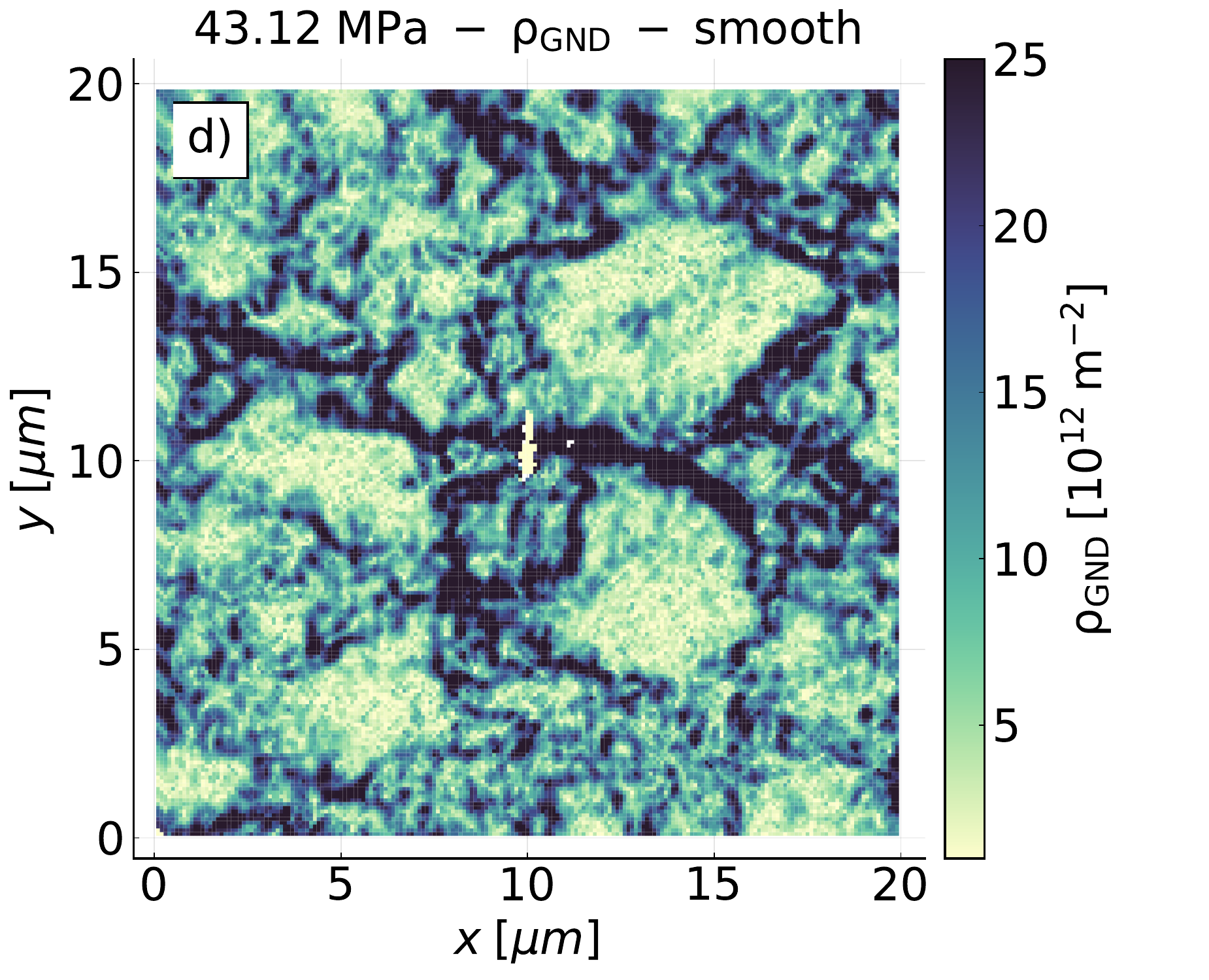}
	\caption{Example for $\mathrm{\alpha_{yz}}$ maps and GND density maps obtained with and without smoothing for the sample compressed up to 43.12 MPa. a) $\alpha_{yz}$ map without smoothing, b) with smoothing, c) GND map without smoothing, d)  with smoothing.}
	\label{GNDComp}
\end{figure}

\begin{figure}[H]
	\centering
		\includegraphics[width=0.24\textwidth]{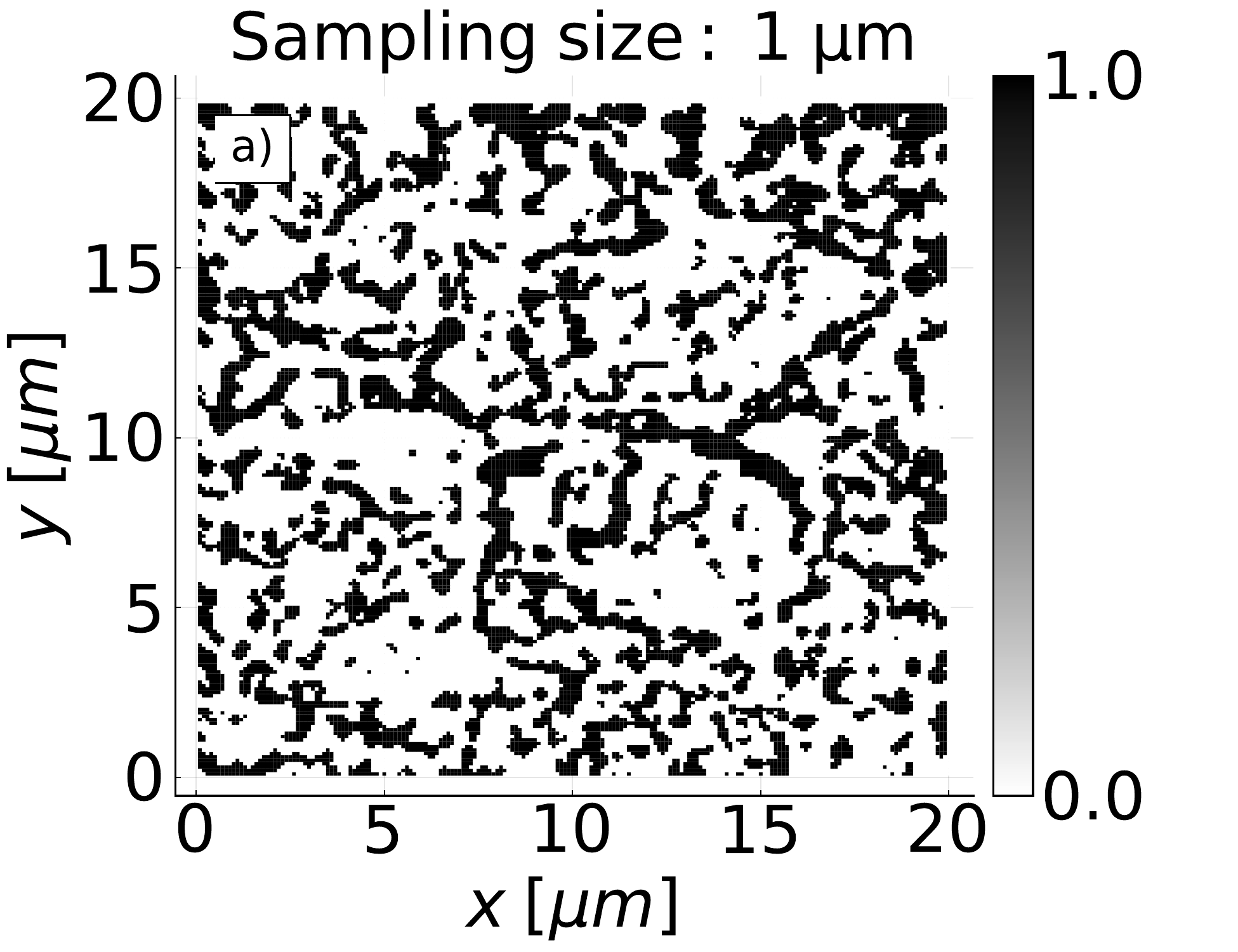}
		\includegraphics[width=0.24\textwidth]{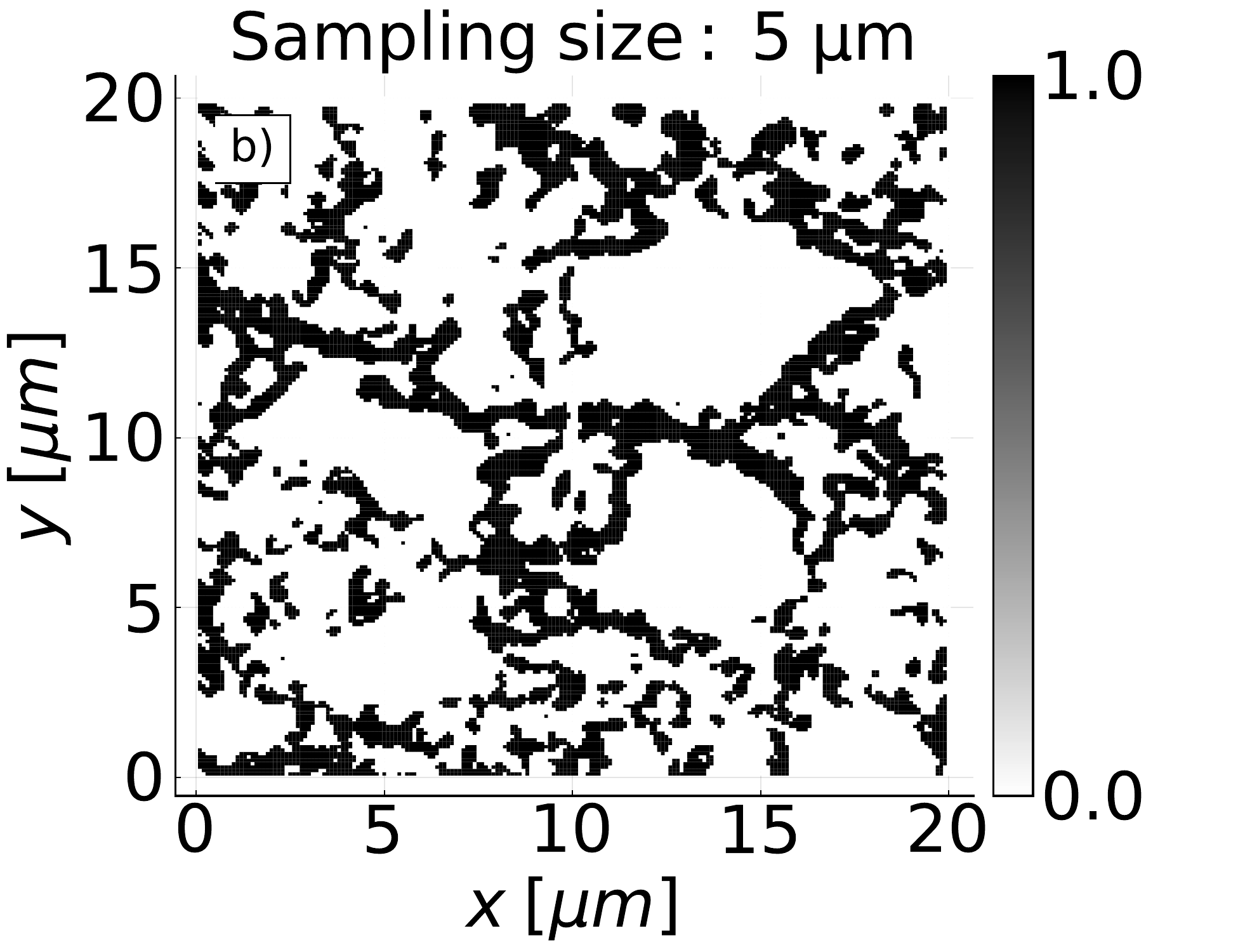} 
		\includegraphics[width=0.24\textwidth]{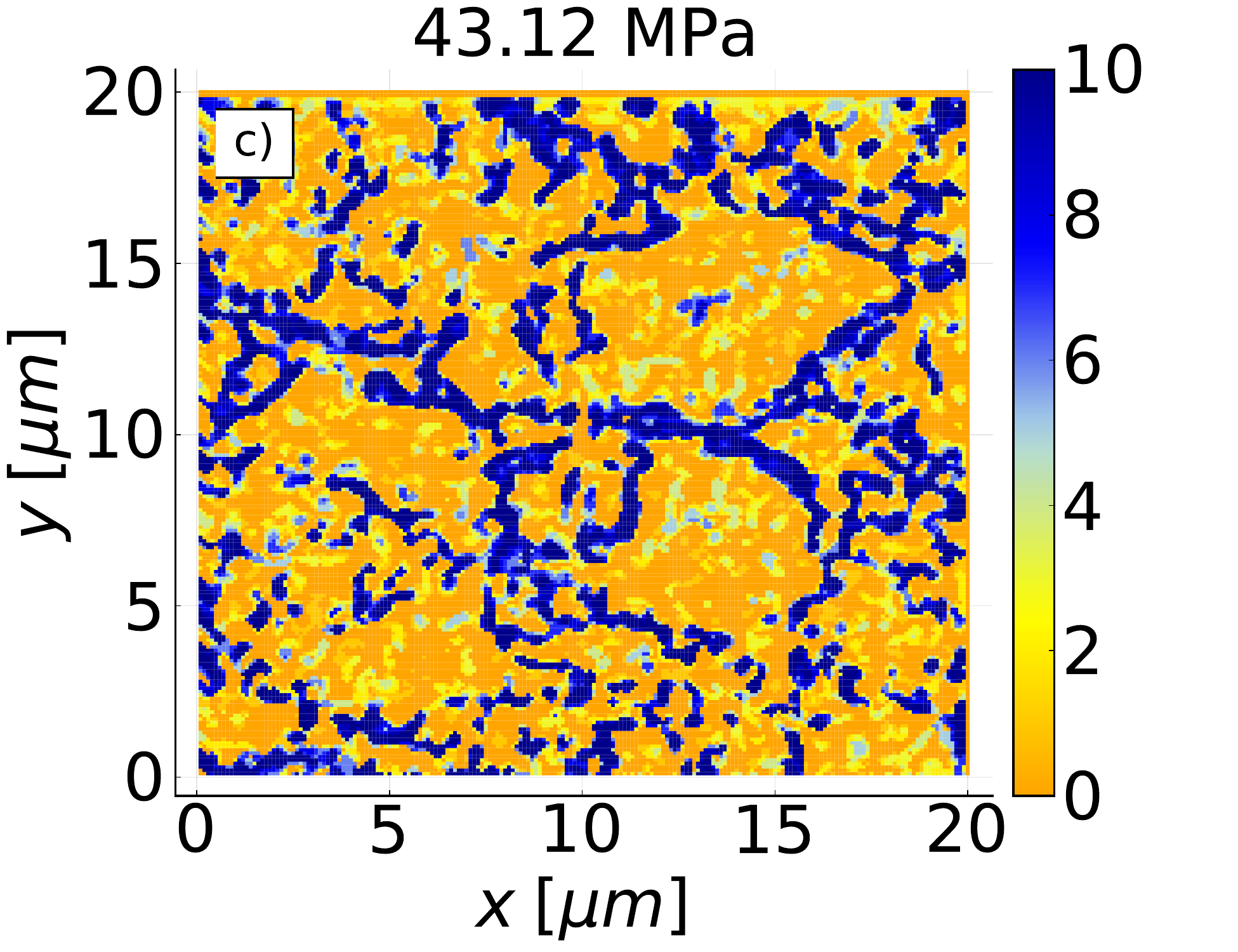}
		\includegraphics[width=0.24\textwidth]{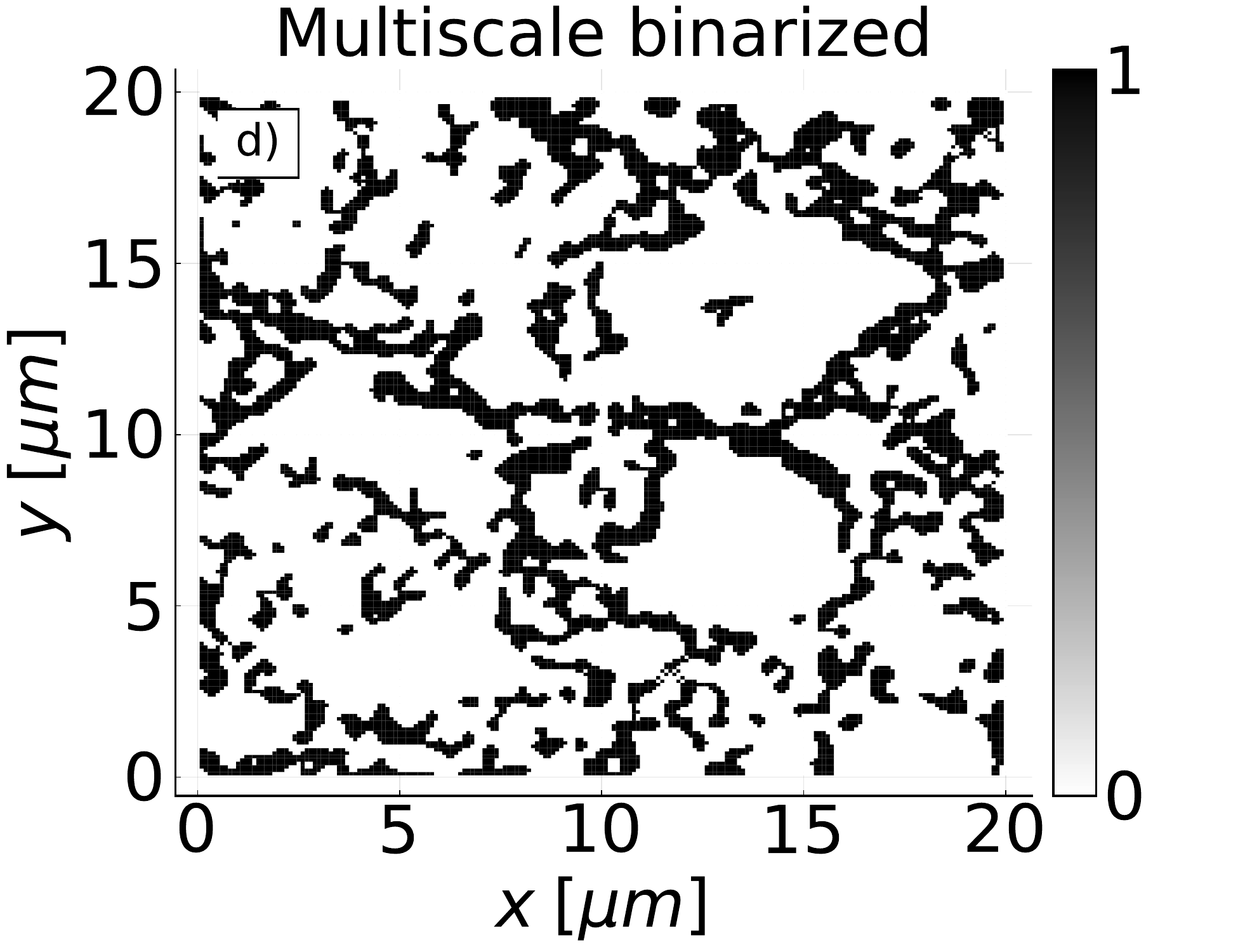} 
		\caption{a) Otsu's binarization method  with box size of 1 $\mathrm{\mu}$m, b) 5 $\mathrm{\mu}$m, c) the added binary maps at all binarization sizes, and d) the map after the final binarization.}
	\label{Otsus}
\end{figure}

One may observe a fluctuating salt-and-pepper like noise on the Nye-tensor component maps (Fig.~\ref{GF5}) and GND density maps (Fig.~\ref{GNDmaps}). This prevents the direct applicability of Otsu's method. In order to eliminate this noise, a smoothing window  was applied to the measurable Nye-tensor components. The maps were convoluted with a circular averaging window of radius $\mathrm{r=150}$ nm. The application of a smoothing window results in a more pronounced dislocation wall structure (Fig.~\ref{GNDComp}).
A globally applied binarization method discussed above may ignore those dislocation walls, which may have a lower dislocation density  than the thickest dislocation ensembles. In order to avoid this  problem a multiscale binarization method was developed. The area map was subdivided  into squared sub-areas and Otsu's method was applied separately for each sub-areas (Fig.~\ref{Otsus}). By repeating this algorithm with areas  with different sizes and by adding up the maps binarized with different scales we could obtain a purely bimodal histogram for the image (Fig.~\ref{Otsus}). Those pixels were considered as dislocation walls which had a higher value than the intensity value corresponding to the minimum of the histogram valley. This method is a powerful tool to obtain not only the global, but also the globally invisible, locally present dislocation walls (see Fig.~\ref{Otsus}).

\subsection{Burgers vector analysis}
\label{sec:bv_analysis}
As it was discussed above only  the $\alpha_{iz},  \; i = x,y,z;$ components of the Nye-tensor can be determined from a HR-EBSD measurement without any further assumption regarding the dislocation system (Fig.~\ref{GF5}).  Therefore, according to Eq.~(\ref{eq_alpha}) the vector constructed from the available Nye-tensor components
\begin{equation}
    \vec{B} = (\alpha_{xz}, \alpha_{yz}, \alpha_{zz})
\end{equation}
is 
\begin{equation}
 B_i=\sum_t b_i^t \rho^t \cos(\vartheta^t) 
\end{equation}
where $\vartheta^t$ is the angle between the line direction of the $t^{\mathrm{th}}$ type of dislocation and the surface normal vector. To characterize the type and sign of the dislocation at a given point of the scanned surface the method introduced in \cite{zoller2020microstructure} is followed, that is, the quantity 
\begin{equation}
    a_i = \cos \left(\varphi_i \right) = \frac{\vec{B} \cdot \vec{b}_i}{B b_i} \label{defa}
\end{equation}
can be calculated where the index $i$ goes through all the 6 Burgers vectors existing in FCC crystals \cite{zoller2020microstructure}. Certainly one cannot determine the relative population of the different type of dislocations from $\vec{B}$, but according to the definition given by Eq.~(\ref{defa}) if the $\rho^t$ density of one of the Burgers vectors is dominantly larger than the other ones, the absolute values of the corresponding $a_i$ are close to 1. Therefore, $a_i$ values can help to describe the type of dislocations at the sample surface. To visualize this the product of the $a_i$ and GND maps were calculated for the 6 possible Burgers vectors. (Typical results can be seen in Fig.~\ref{BurgEval}.)

\section{Results and discussion}

As a first step X-ray line profile measurements with $\{020\}$ Bragg reflection were performed on the $(010)$ surface of the undeformed and the 6 samples deformed up to different stress levels. According to earlier investigations on deformed Cu single crystals oriented for ideal multiple slip \cite{Ungar:gk0172} for this reflection $\Lambda=0.783$. The intensity distributions were analysed with the restricted momentum method explained earlier. Both the $2^{\mathrm{nd}}$ and the $4^{\mathrm{th}}$ order restricted moments were evaluated. 

\begin{figure}[H]
	\centering
		\includegraphics[width=0.7\textwidth]{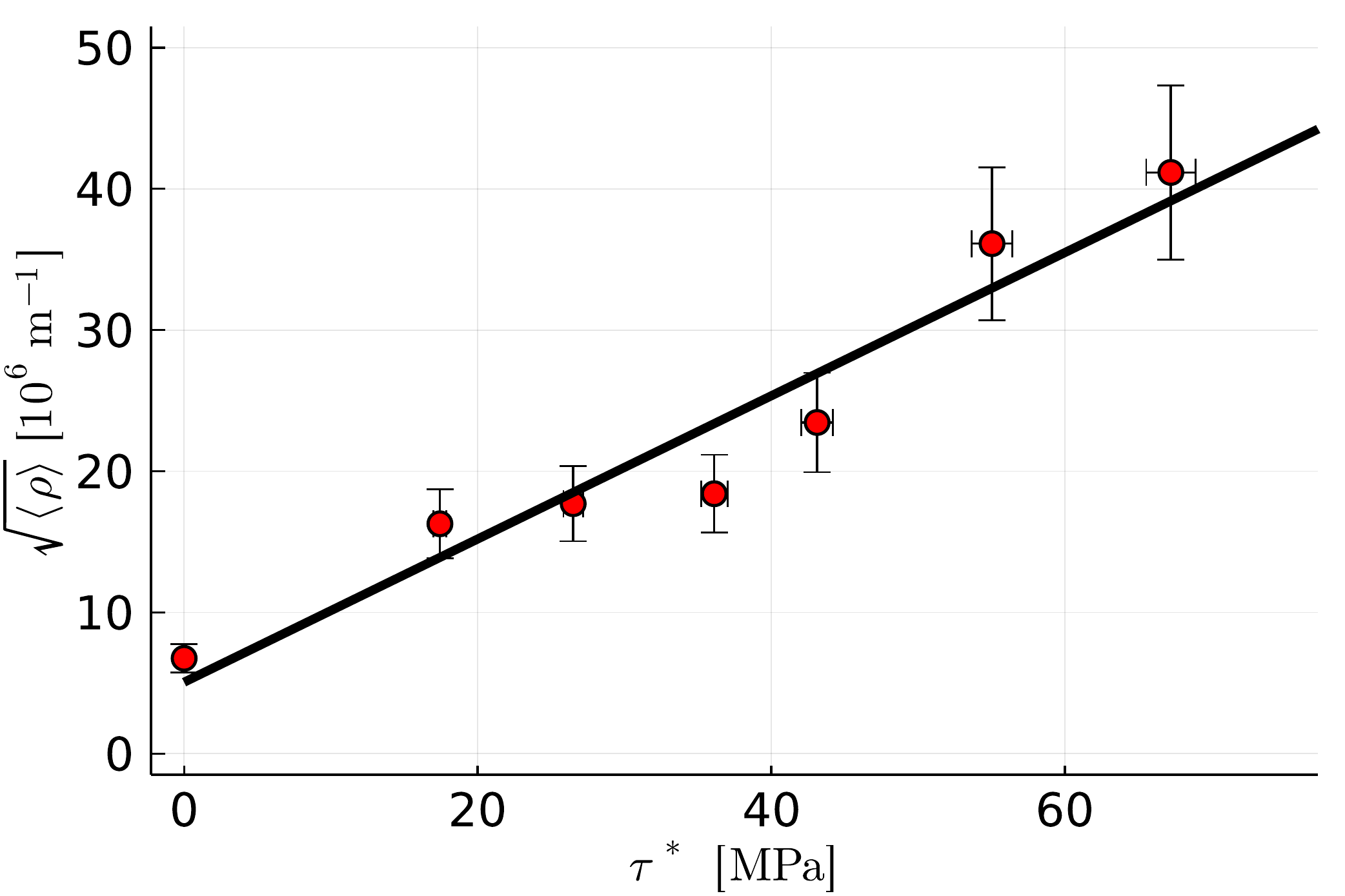}
	\caption{The $\sqrt{\langle\rho\rangle} - \tau^*$ relation, where $\langle\rho\rangle$ is the average dislocation density measured by X-ray and $\tau^*$ is the resolved shear stress.}
	\label{SqrtRhoTau}
\end{figure}

As it can be seen in Fig.~\ref{SqrtRhoTau} the  Taylor linear relation between the square root of the dislocation density and the resolved shear stress is fulfilled. The small deviation seen at $\tau^*=26.5$ MPa and $\tau^*=36.11$ MPa stress levels can be attributed to the fact that close to the end of stage II the dislocation population may differ from the one obtained in the investigations presented in \cite{Ungar:gk0172}. As a consequence the contrast factor can be slightly different from the one used. 
The results are in agreement with the earlier investigations of Székely et al.~\cite{szekely2001statistic}. It should be noted, however, that the X-ray detector used in the measurements reported here has a much better signal-to-noise ratio than the one used earlier resulting in an improved accuracy of the current study. 

\begin{figure}[H]
	\centering
		\includegraphics[width=0.7\textwidth]{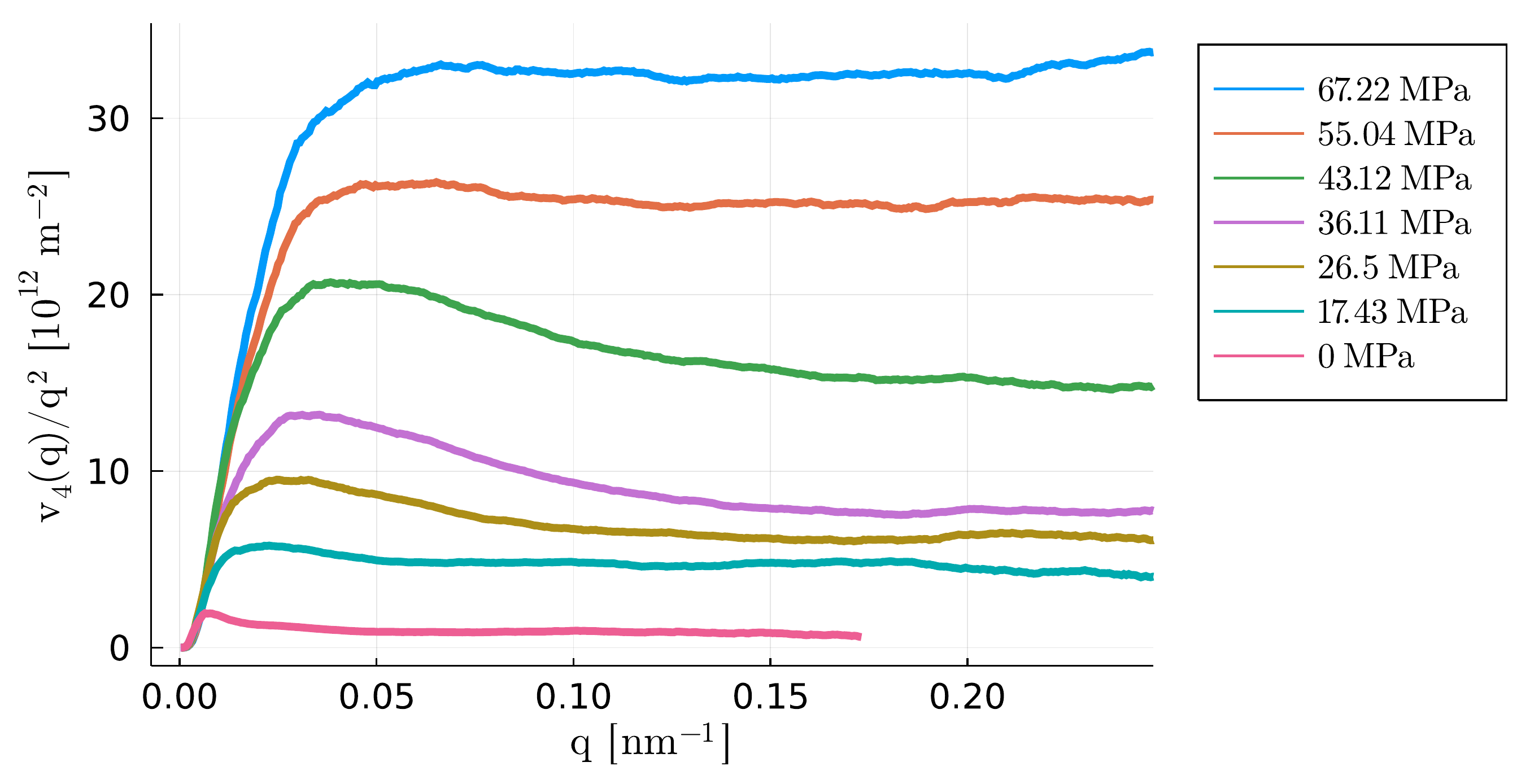}
	\caption{The $v^4(q)/q^2$ versus $q$ curves at different compression levels. The corresponding resolved shear stresses are indicated in the upper right corner.}
	\label{M4Moments}
\end{figure}

In Fig.~\ref{M4Moments} the $v^4(q)/q^2$  restricted moments are plotted for the 7 samples. It can be seen even without any curve fitting that the asymptotic part of the curves tend to a constant value that increases monotonically with the applied stress. Since the asymptotic value of the $v^4(q)/q^2$ is proportional to the average dislocation density this is in accordance with the results discussed above. It is remarkable, however, that the maximum values of the curves normalized with the asymptotic value is not monotonous with the stress. It has a clear maximum at 36.11 MPa stress level. After performing the fitting of the function given by Eq.~(\ref{eqf}) the $\sigma$ value defined by Eq.~(\ref{avg.dis.dens.fluct.}) can be determined.      

\begin{figure}[H]
	\centering
		\includegraphics[width=0.7\textwidth]{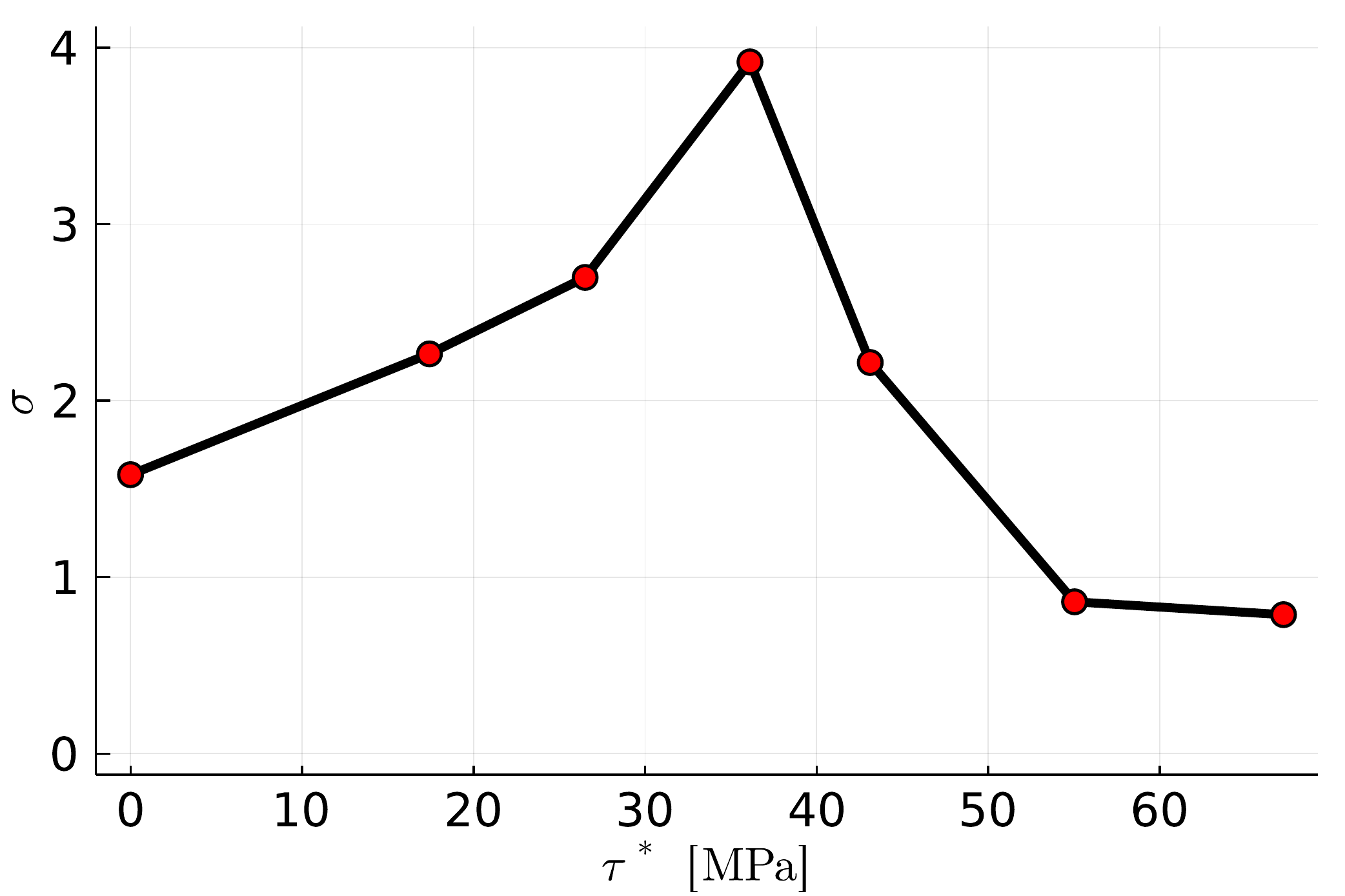}
	\caption{The $\sigma(\tau^*)$ function is represented, where $\sigma$ is the average dislocation density fluctuation.}
	\label{FluctStress}
\end{figure}

In agreement with the ``phenomenological'' feature mentioned above the $\sigma$ versus $\tau$ curve exhibits a sharp maximum at $\tau=36.11$ MPa (see Fig.~\ref{FluctStress}) corresponding to the stage II to stage III transition stress level (see Fig.~\ref{StressStrainCurve}). The results obtained are in agreement with the ones reported earlier on the same material \cite{szekely2001statistic}.

\begin{figure}[H]
	\centering
		\includegraphics[width=0.7\textwidth]{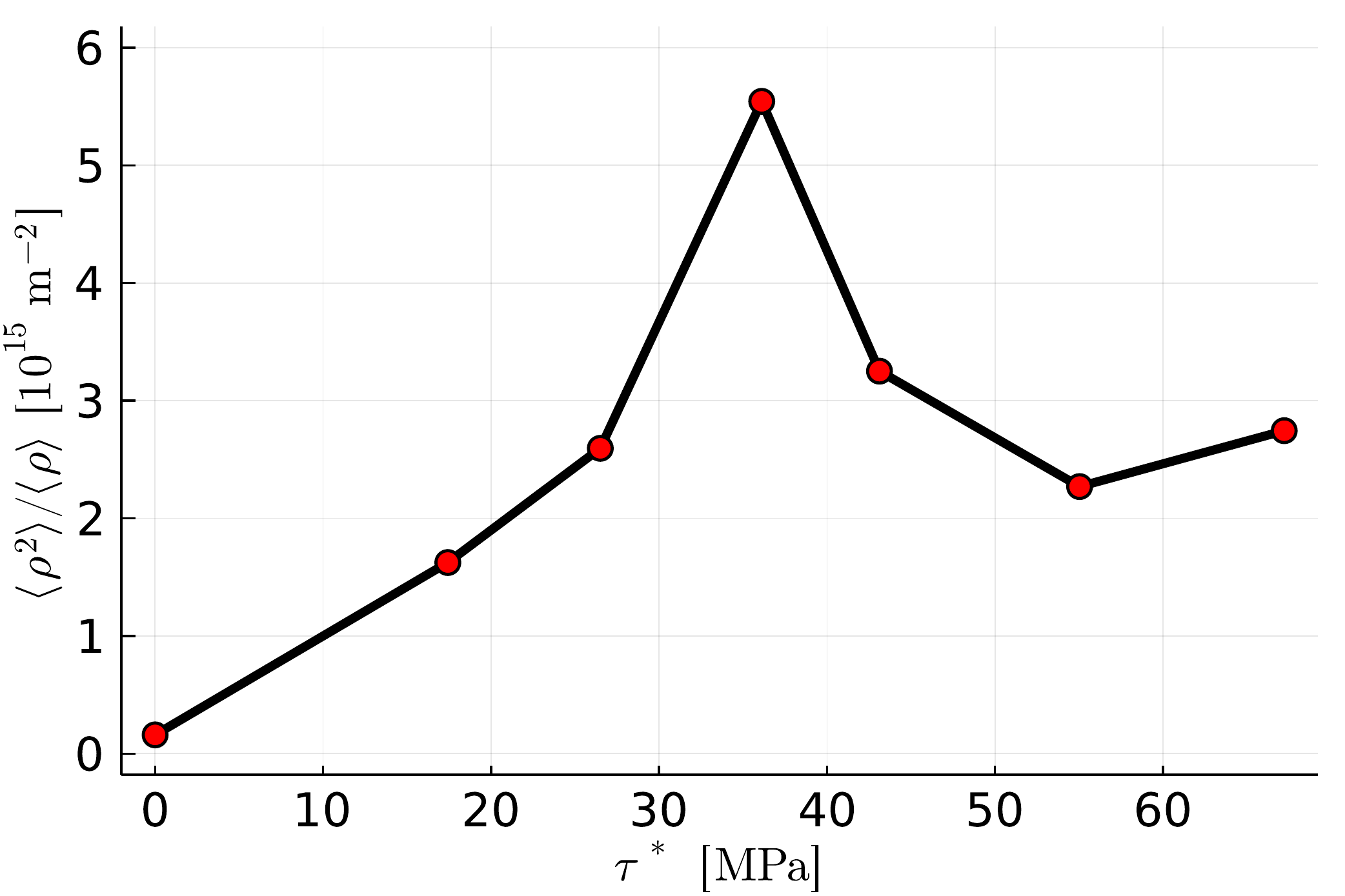}
	\caption{The ``apparent'' dislocation density $\rho_w^{app}$ as a function of the resolved shear stress.}
	\label{rhowStress}
\end{figure}

As it was suggested earlier by Mughrabi et al.~\cite{mughrabi2002long,ungar1984x} the dislocation system can be envisaged as a composite of ``hard'' dislocation walls with dislocation density of $\rho_w$ and ``soft'' cell interiors with dislocation density $\rho_c$. Within this model 
\begin{equation}
 \langle \rho \rangle=f \rho_w +(1-f) \rho_c
\end{equation}
and
\begin{equation}
 \langle \rho^2 \rangle=f \rho_w^2 +(1-f) \rho_c^2
\end{equation}
where $f$ is the volume fraction of the cell walls.
Since according to earlier investigations \cite{mughrabi2002long,ungar1984x} $f$ is in the order of 0.1, and $\rho_w$ is an order of magnitude higher than $\rho_c$ the second term in $ \langle \rho^2 \rangle$ can be neglected so
\begin{equation}
 \langle \rho^2 \rangle\approx f \rho_w^2
\end{equation}
With this the quantity $\rho_w^{app}=\langle \rho^2 \rangle/\langle \rho \rangle$, that can be determined directly from the X-ray line profile, is
\begin{equation}
 \rho_w^{app}=\rho_w \frac{1}{1+\frac{(1-f)\rho_c}{f \rho_w}}.
\end{equation}
If $\rho_c$ is small then the ``apparent'' dislocation density $\rho_w^{app}\approx\rho_w $. According to Fig.~\ref{rhowStress} in stage II, $\rho_w^{app}\approx\rho_w $  increases monotonically and at the stage II to III transition stress level it has a maximum. In stage III at large enough stress it tends to saturate.

Based on the X-ray line profile results it can be concluded that during stage II the dislocation distribution becomes more and more inhomogeneous within the studied deformation range, as dense dislocation walls are formed. At a given deformation level, however, the dislocation density in the walls reaches a maximum level, dislocation annihilation prevents the further increase of the dislocation density. This process is called dynamic recovery \cite{mughrabi2002long}. In stage III new walls and an increase of the dislocation density in the cell interiors is needed to accumulate more dislocations. According to Fig.~\ref{rhowStress} for large enough stress levels the term $(1-f)\rho_c/f/\rho_w$ is in the order of unity.   

As it is seen above X-ray line profile analysis is a rather powerful method to determine some average statistical properties of the dislocation microstructure, but certainly it is not able to say anything about the actual dislocation morphology.

\begin{figure}[H]
  \centering
  \includegraphics[scale=.22]{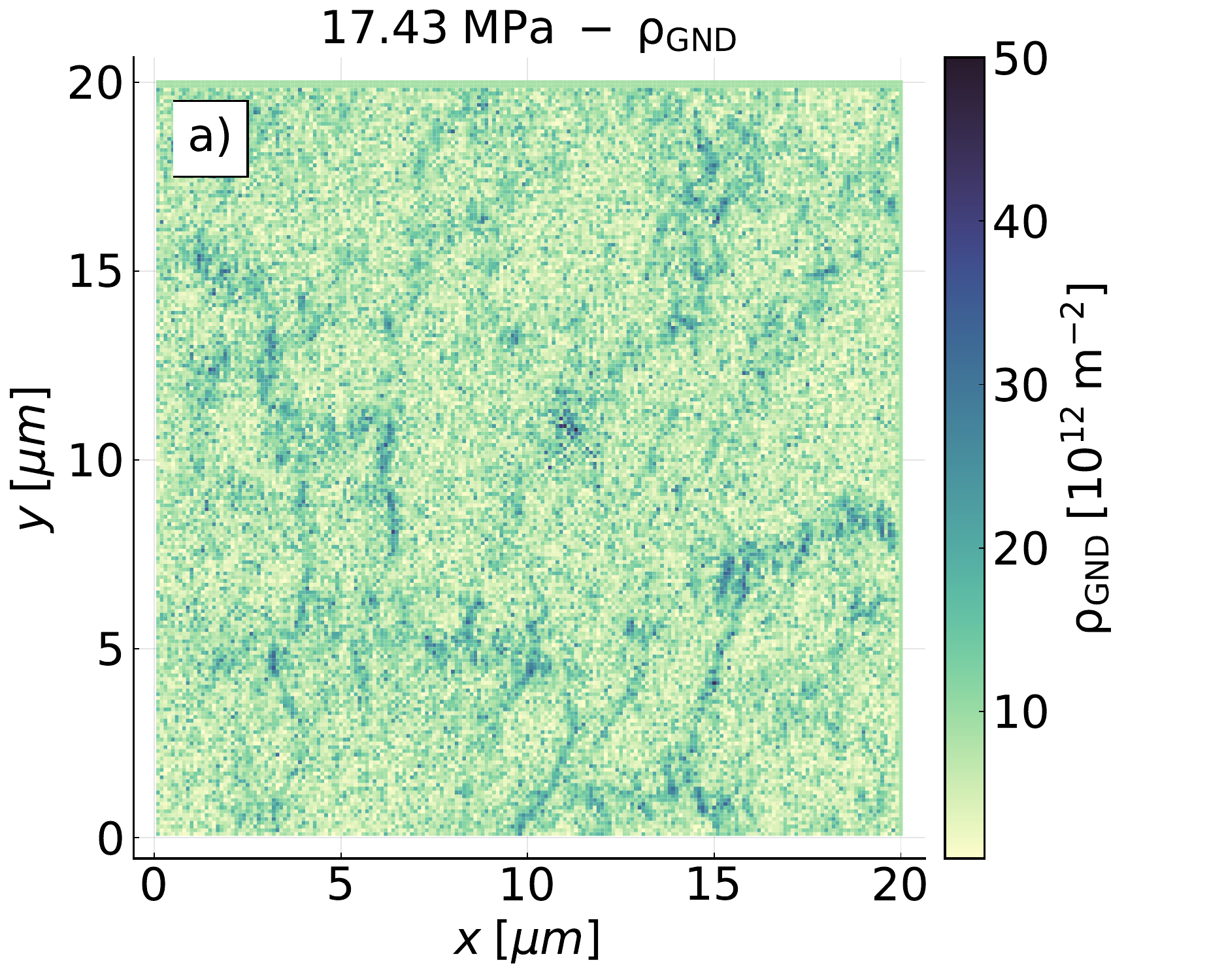} 
  \includegraphics[scale=.22]{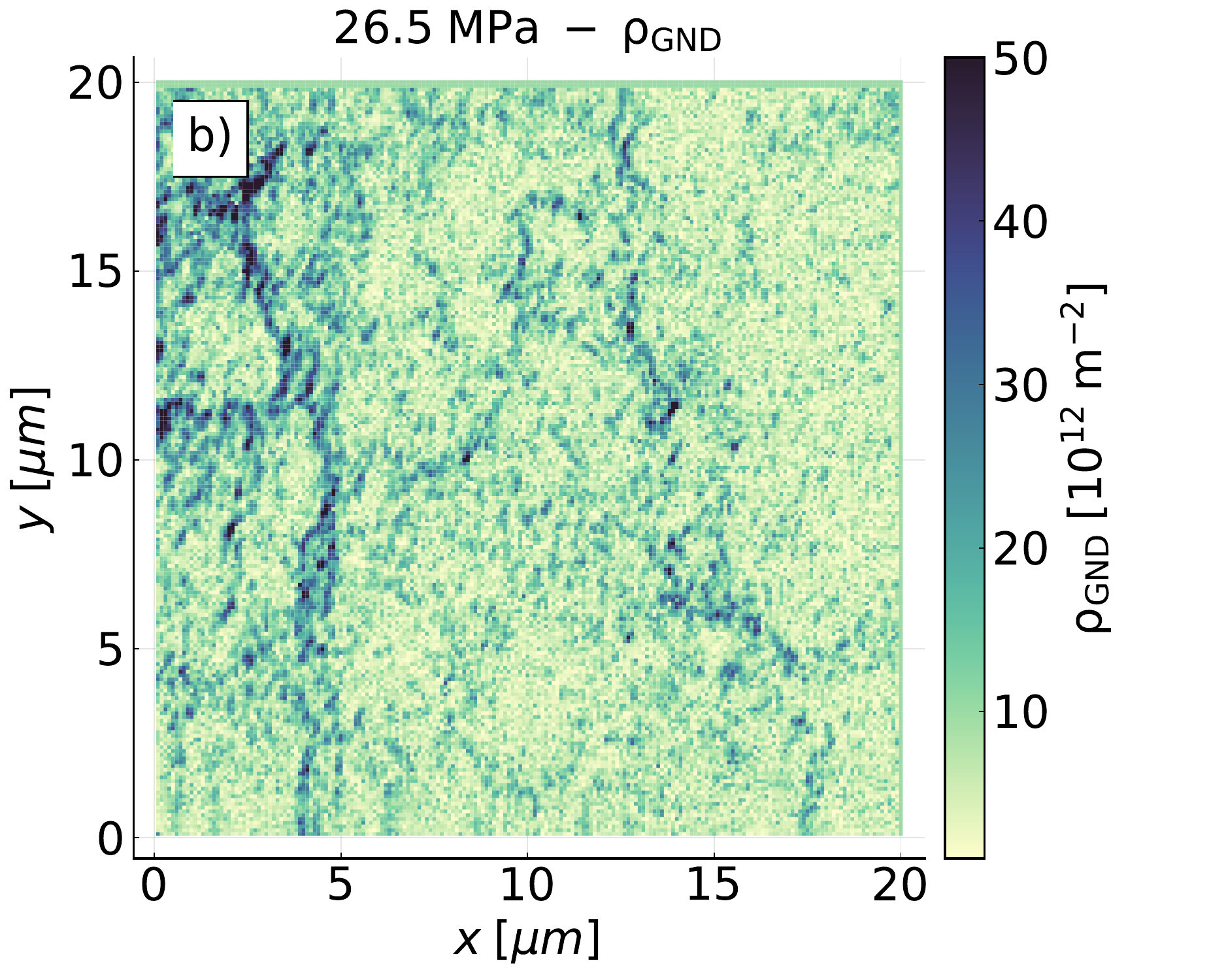} 
  \includegraphics[scale=.22]{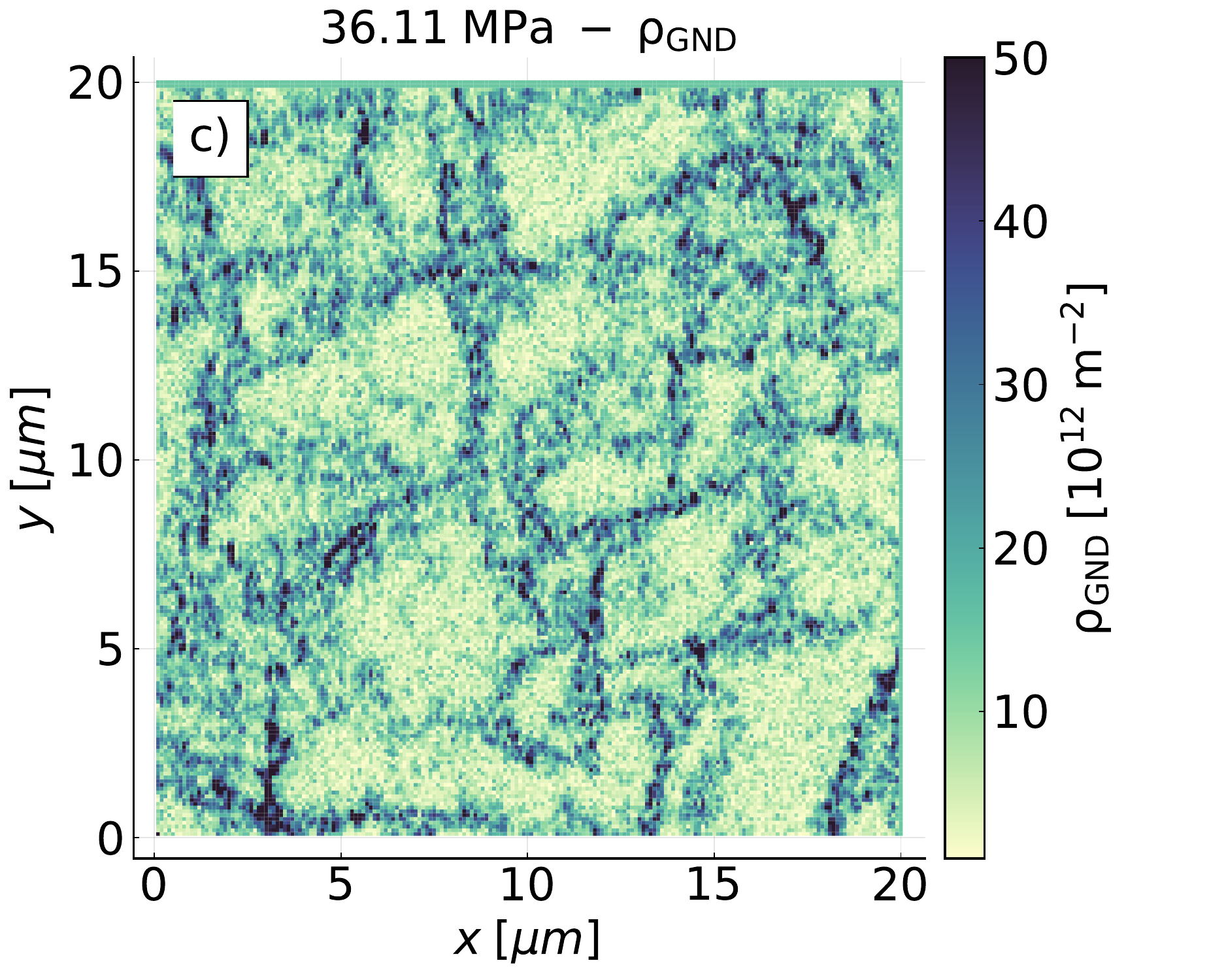} 
  \includegraphics[scale=.22]{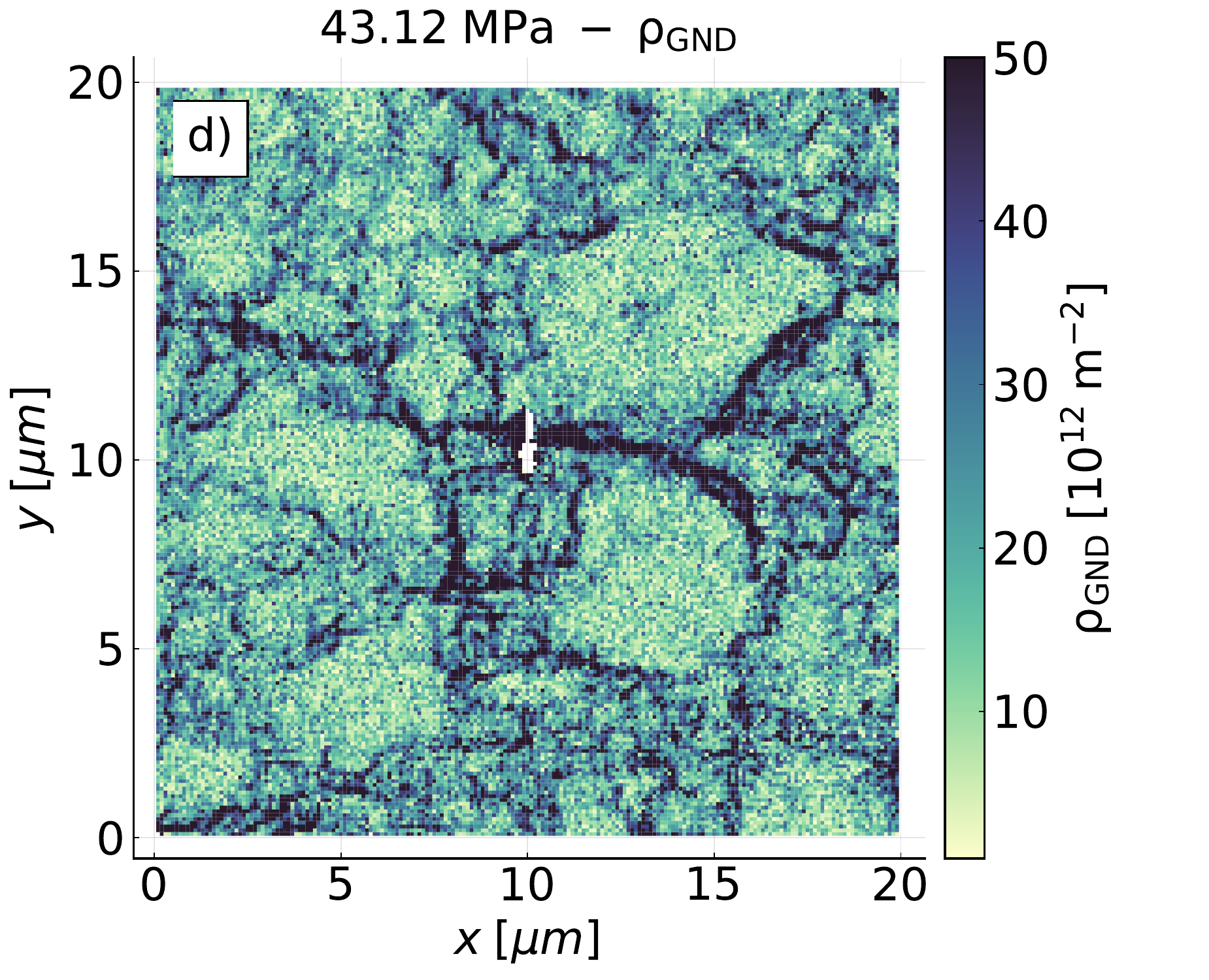} 
  \includegraphics[scale=.22]{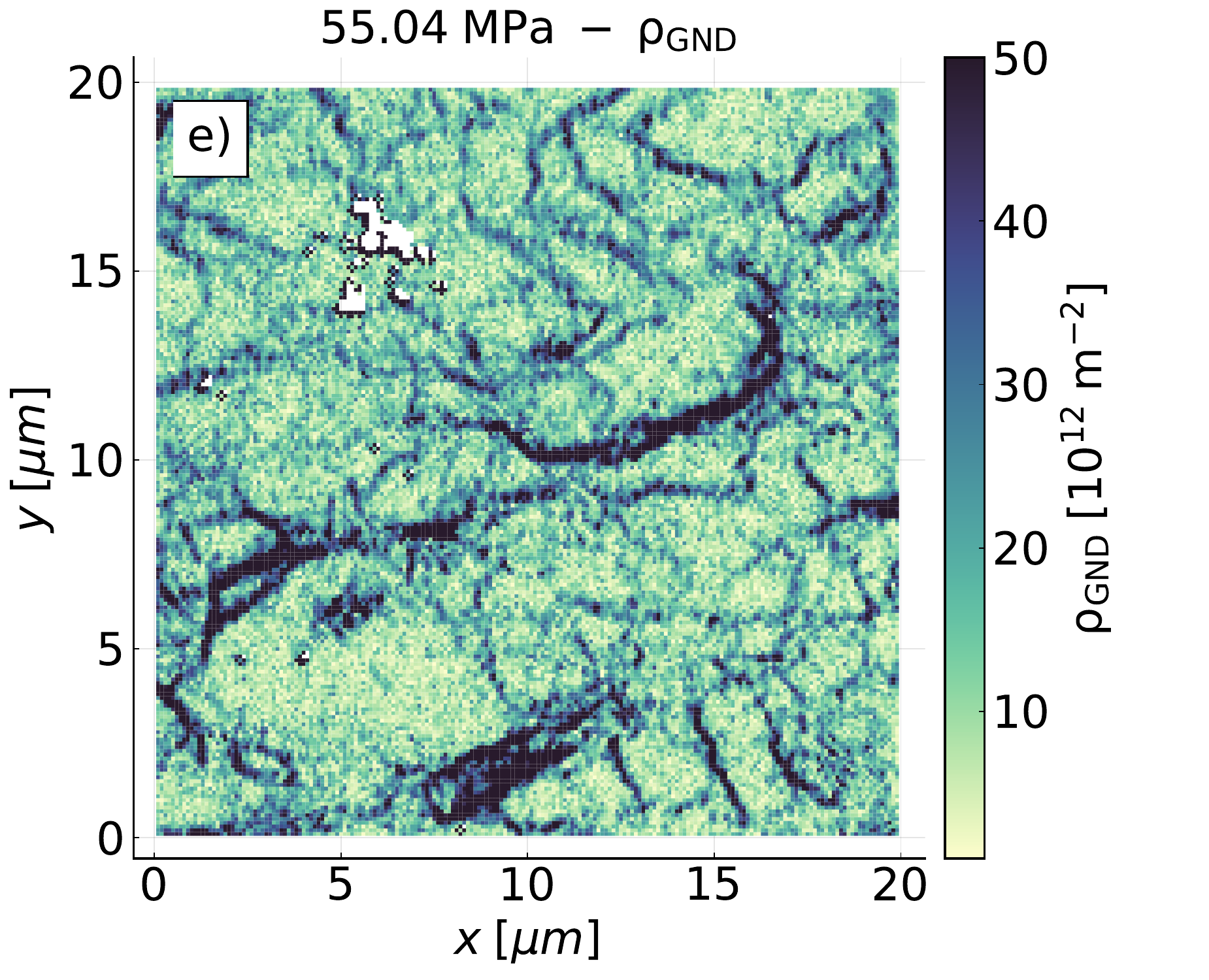} 
  \includegraphics[scale=.22]{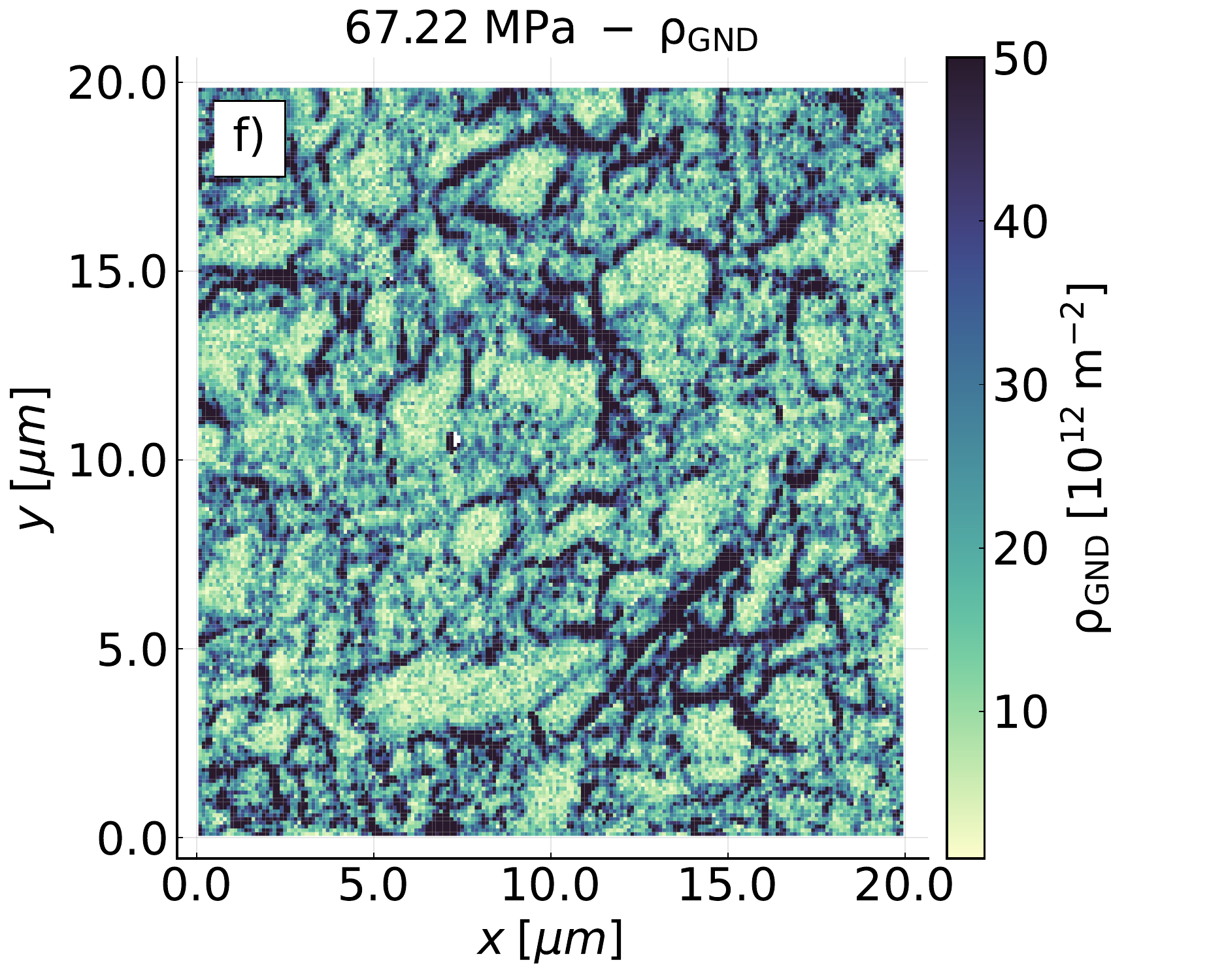}
	\caption{The GND density maps obtained on samples deformed up to (a) 17.43, (b) 26.5, (c) 36.11, (d) 43.12, (e) 55.04, and (f) 67.22 MPa. }
	\label{GNDmaps}
\end{figure}

\begin{figure}[H]
  \centering
        \includegraphics[scale=.22]{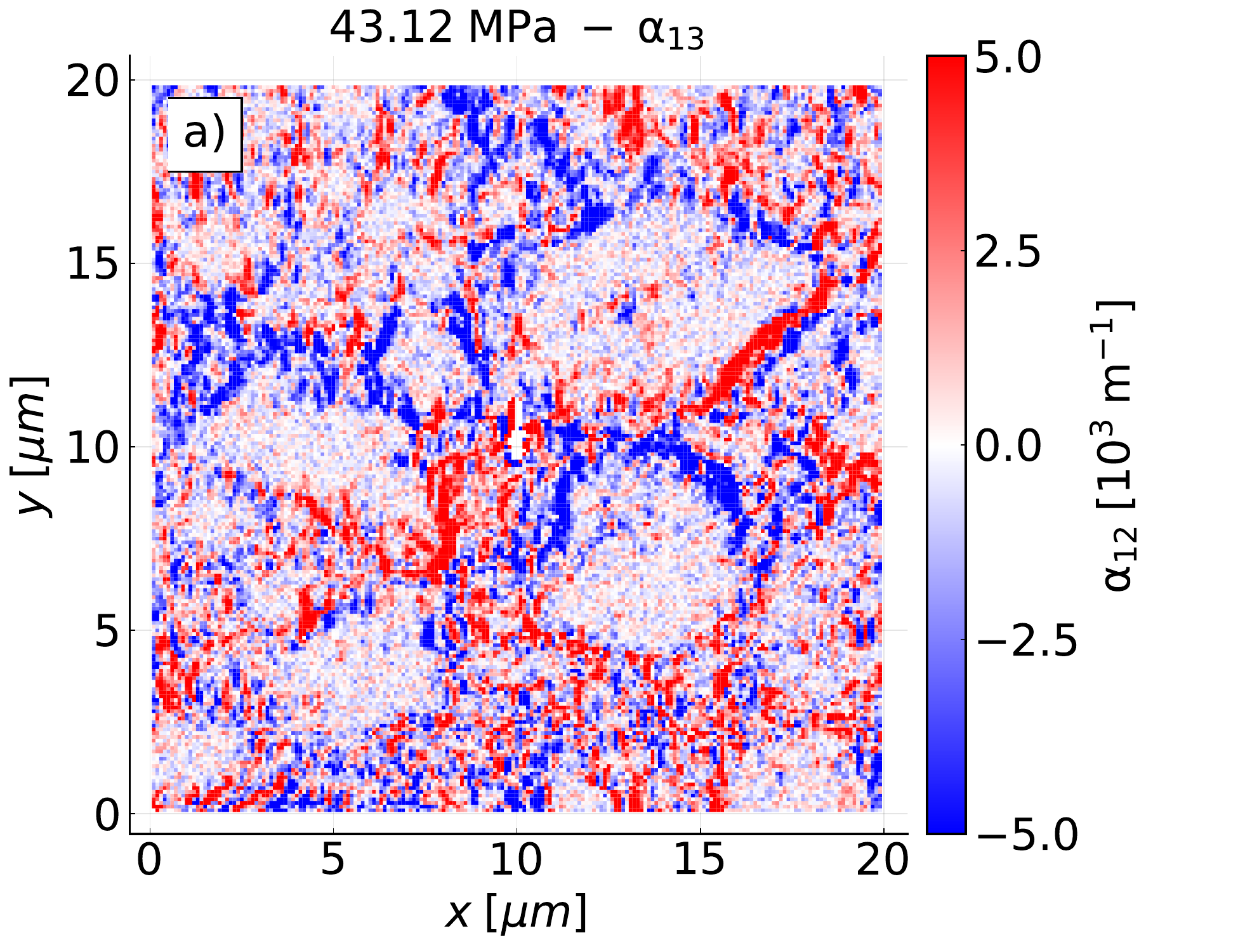} 
        \includegraphics[scale=.22]{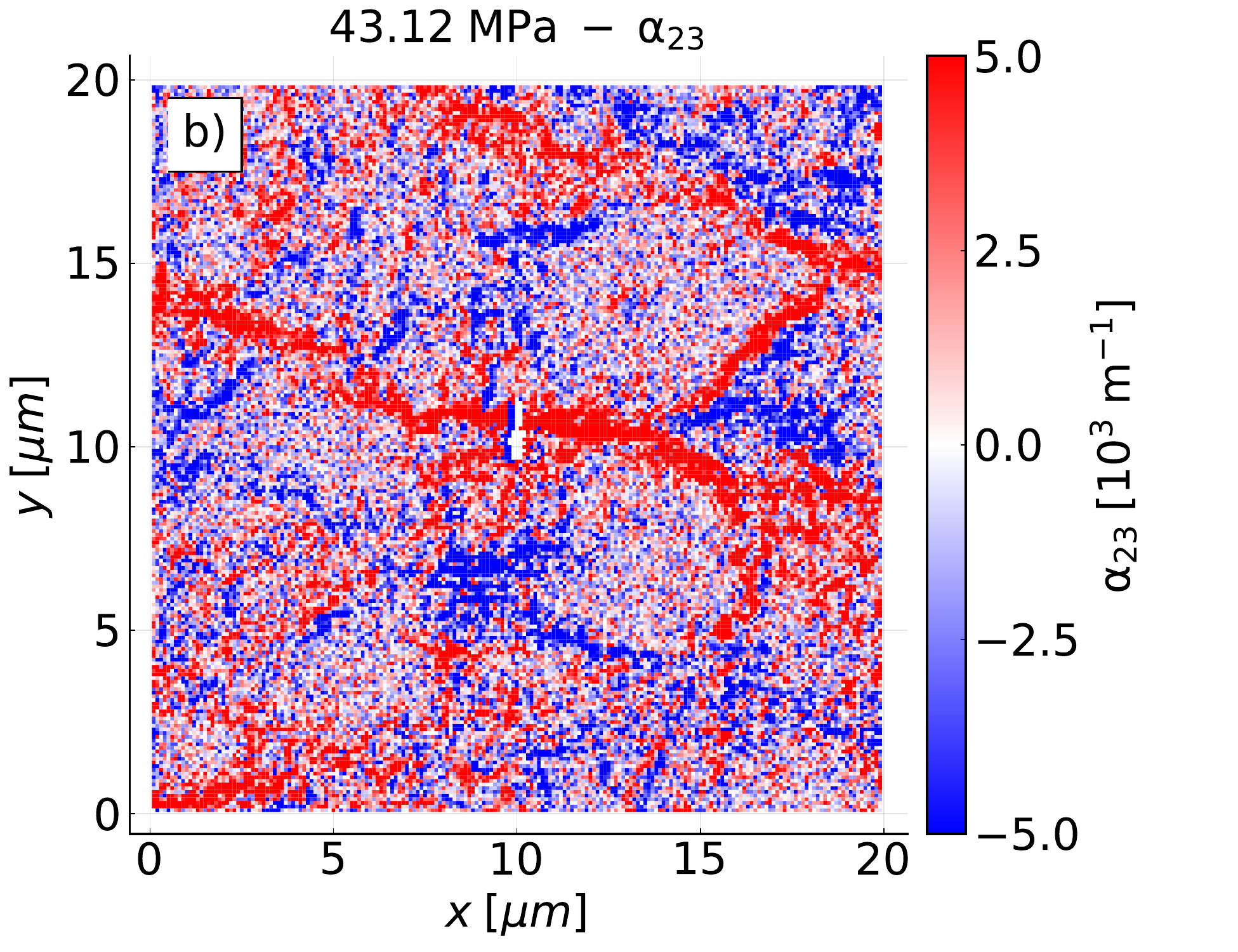} 
        \includegraphics[scale=.22]{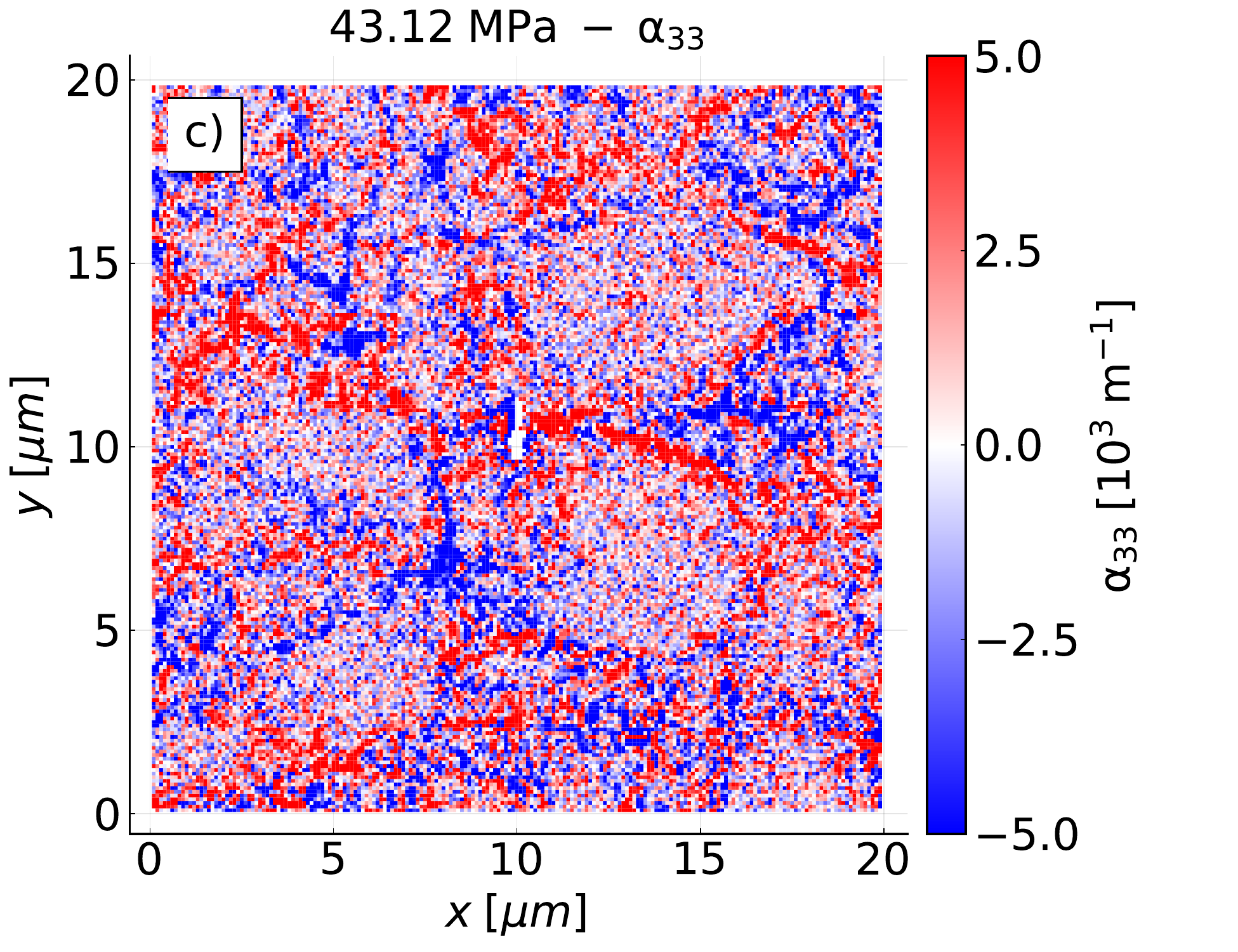} 
        \includegraphics[scale=.22]{Fig13_d.pdf} 
		\includegraphics[scale=.22]{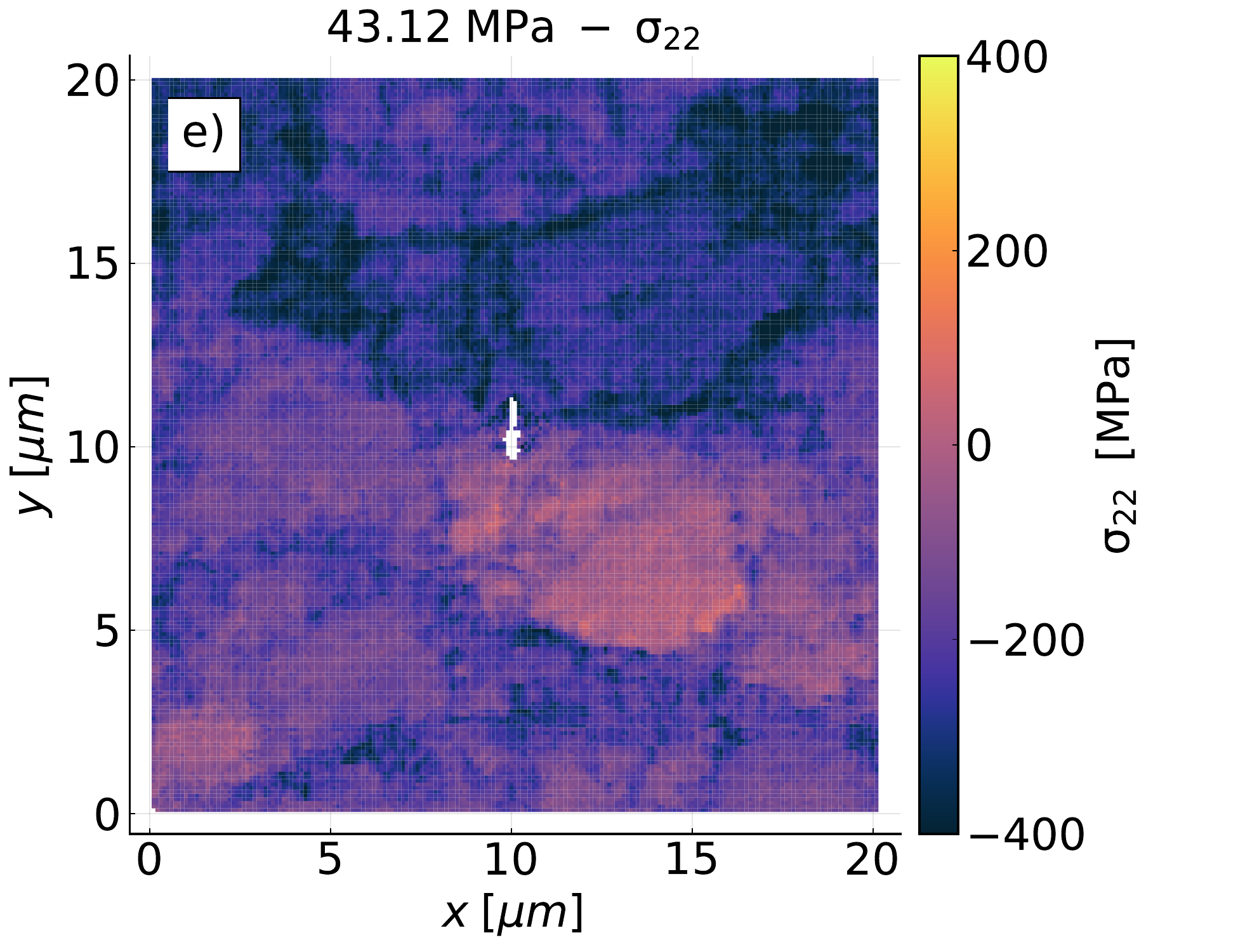}
		\includegraphics[scale=.22]{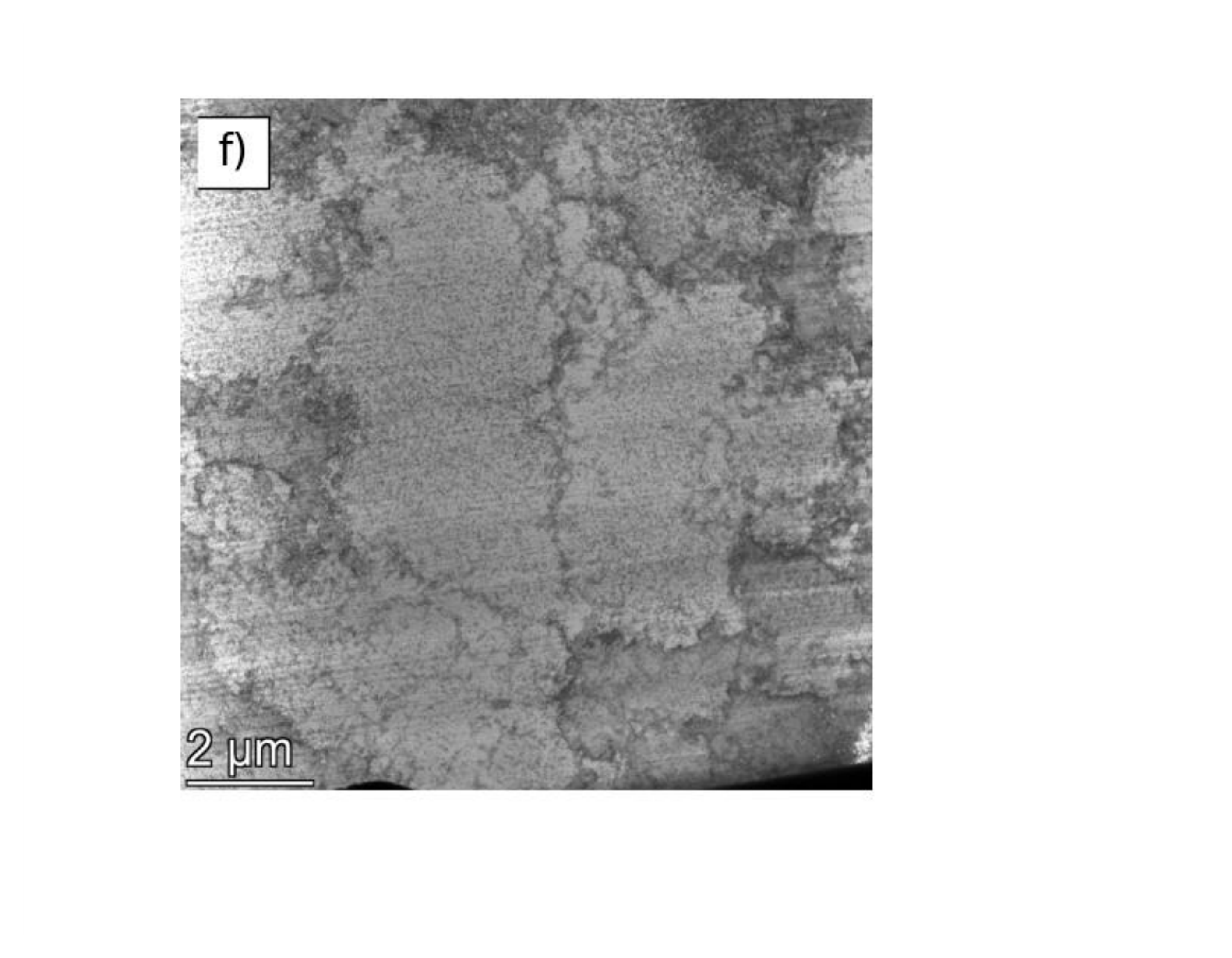}
	\caption{The maps of the (a) $\alpha_{13}$, (b) $\alpha_{23}$, and  (c) $\alpha_{33}$ components, (d) the GND, the (e) $\sigma_{22}$, and (f) a TEM picture obtained on the sample deformed up to 43.12 MPa. Notice that the scale and the observation site on the TEM picture is different than on the other ones.}
	\label{GF5}
\end{figure}

Beside the traditionally applied TEM  \cite{oudriss2016length} the relatively recently developed HR-EBSD method offers a new perspective to directly study the dislocation microstructure. A big advantage of the HR-EBSD is that a much larger area can be studied than by TEM. Moreover, the sample preparation is much easier. Figure \ref{GNDmaps} shows the GND maps obtained on the 6 deformed samples.  
At each deformation level a clear cell structure can be seen with increasing  volume fraction of the cell walls. In Fig.~\ref{GF5} the maps of the three $\alpha_{iz}$ components, the GND density, the $\sigma_{yy}$ stress, and a TEM picture obtained on the sample deformed is plotted.
Similar picture were obtained at the other stress levels studied.

According to Fig.~\ref{GF5}, as it was assumed earlier \cite{mughrabi2002long,ungar1984x}, long-range internal stress develops in the cell interiors. Unlike X-ray line profile analysis, HR-EBSD is a direct method to determine the local stress state of the sample, so the result obtained is a direct evidence of the presence of long-range internal stresses.

\begin{figure}[H] 
  \centering
  \includegraphics[width=0.7\textwidth]{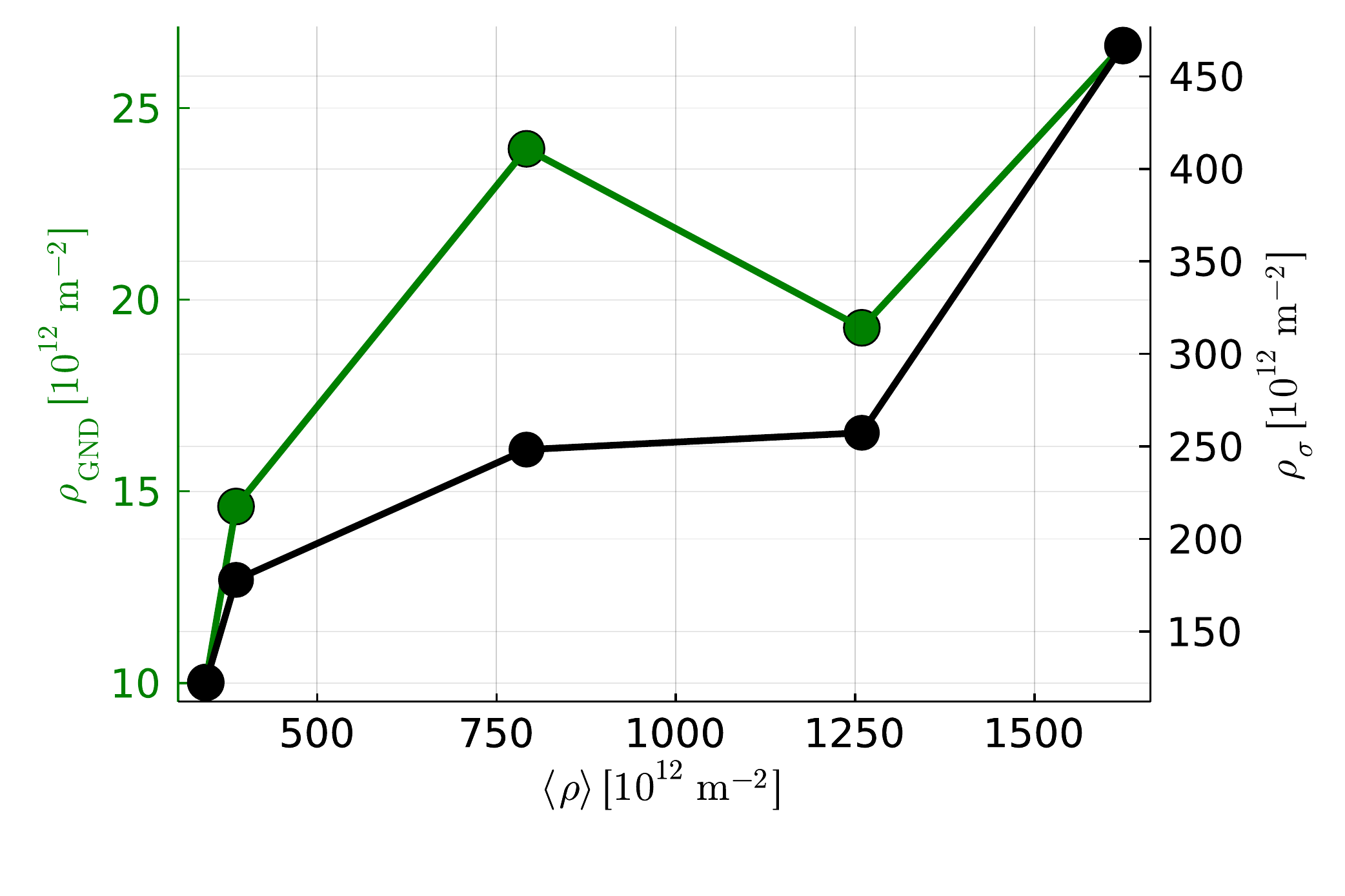}
	\caption{The dislocation densities obtained from the stress probability distribution (black curve) and from the Nye's tensor components (green curve) versus the dislocation density obtained by X-ray line profile analysis. 
	}
	\label{GNDVals}
\end{figure}

The dislocation density was also determined from the stress maps by the restricted moment analysis of the internal stress distribution. In order to reduce the error the average of the $\rho^*_{ij}$ values were calculated for the 5 independent components of the stress tensor. The results obtained are plotted in Fig.~\ref{GNDVals}. As it is seen there is correlation between the $\rho_{\sigma}$ average dislocation density obtained from the stress maps and the $\langle \rho \rangle$ density found by the X-ray line profile analysis, but the relation is clearly not linear. One can also note, that in some points (for example in the fourth), a higher difference is present. This is due to the local nature of the HR-EBSD, but also due to the fact that the measurements were carried out on different samples, not on the same surface and place. Due to these reasons, the first point was eliminated from this figure. Moreover, the $C_{ij}$ geometrical factor may vary with stress and due to the finite volume illuminated by the electrons we cannot detect the small dislocation dipoles (see above). So, the HR-EBSD internal stress analysis is a possible method for the determination of the dislocation density, but the issue requires further investigations to be able to produce dislocation density values with high precision.

The average GND density $\rho_\mathrm{GND}$ defined by Eq.~(\ref{eqGND}) was also determined from the Nye's tensor maps. According to Fig.~\ref{GNDVals}, as a general trend $\rho_\mathrm{GND}$ increases with increasing deformation but due to the large dislocation density fluctuation in order to get more precise GND density values one should perform EBSD measurements on a very large area that was not possible with the setup used. 

After the image binarization with the method explained above the fractal dimensions of the $\rho_\mathrm{GND}$ maps plotted in Fig.~\ref{GNDmaps} were also determined by both the Hausdorff ($D_H$) and the correlation dimension ($D_c$) analysis. Since the maps were binarized, every subset of the map is taken into account with the same weight, so the Hausdorff dimension is equal to the box counting dimension ($D_H = D_B$).
\begin{figure}[H]
	\centering
		\includegraphics[width=0.49\textwidth]{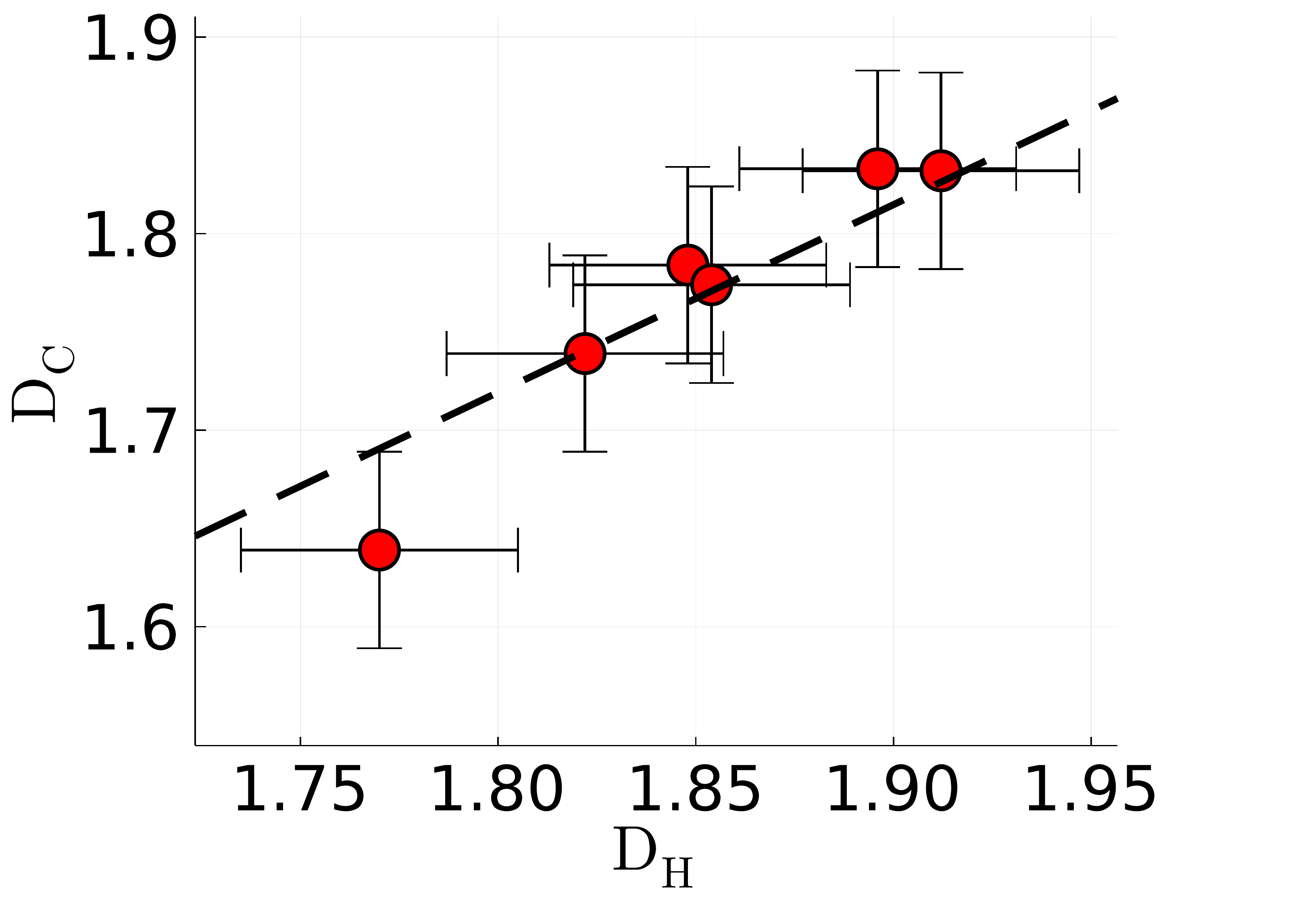}
		\includegraphics[width=0.49\textwidth]{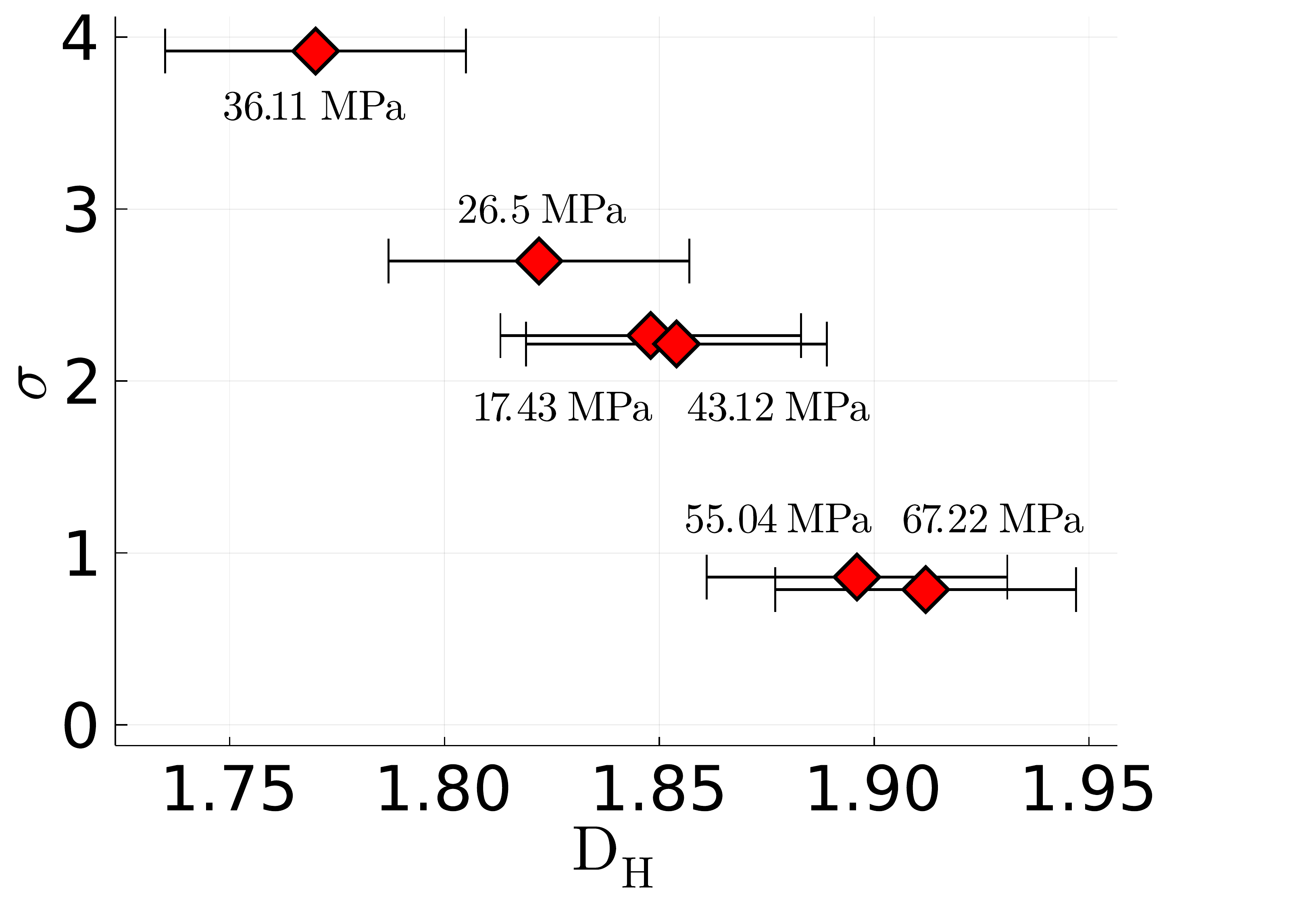}
		\caption{On the left image, the correlation dimension $D_c$ is presented versus the  Hausdorff-dimension $D_H$ obtained by box counting. On the right figure, the average dislocation density fluctuation is shown in function of the Hausdorff-dimension.}
	\label{FractalDim}
\end{figure}
It is found that $D_c\approx 0.95 D_H$ so, the two methods give the same fractal dimension within experimental error. This consistent correlation confirms the formation of a special dislocation structure with non-integer (fractal) dimension. It is, however, a nontrivial result that the fractal dimension is decreasing with increasing relative dislocation density fluctuation $\sigma$. (A similar tendency was reported earlier  \cite{szekely2001statistic}.)

\begin{figure}[H]
  \centering
  \includegraphics[width=0.4\textwidth]{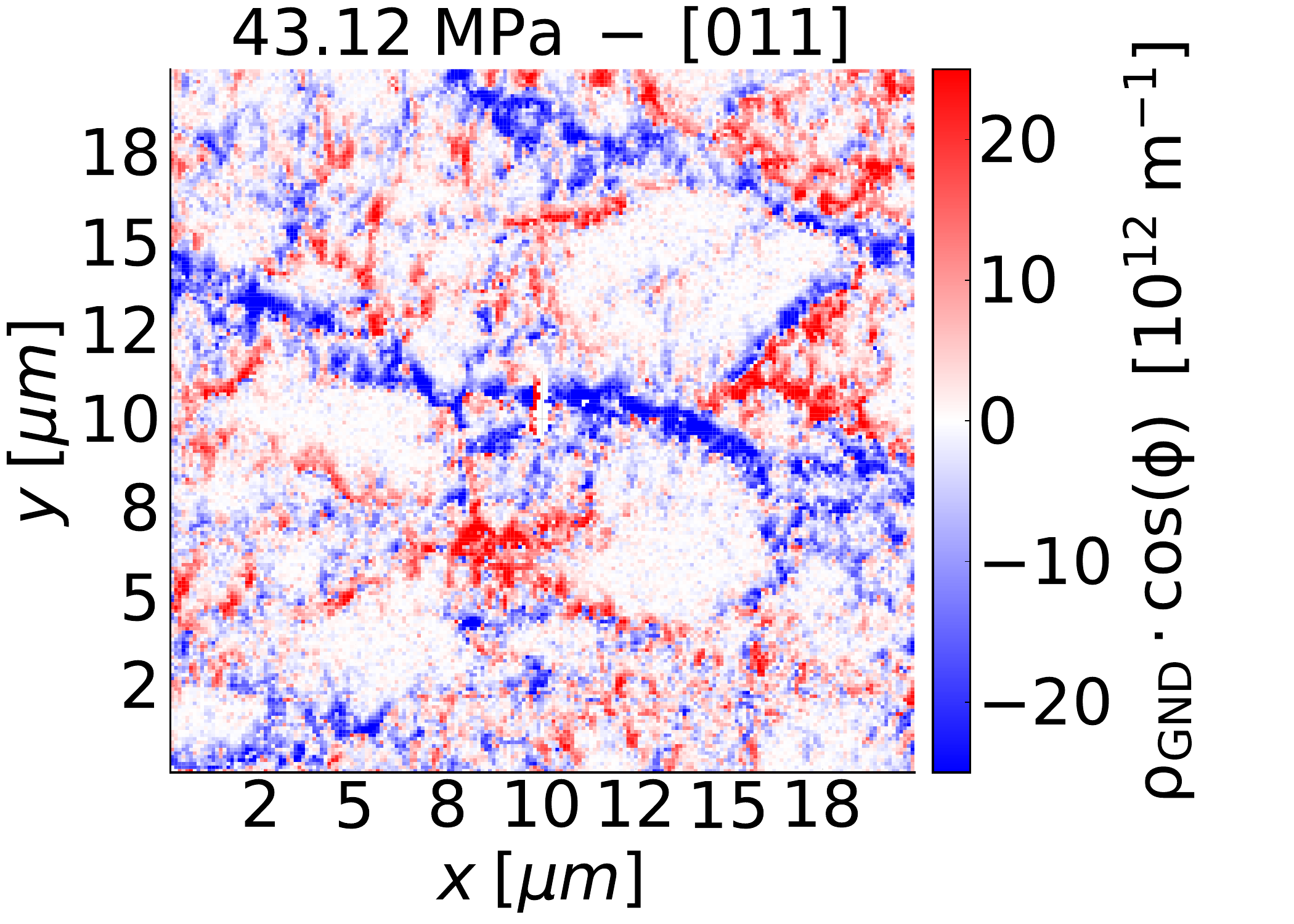}
  \includegraphics[width=0.4\textwidth]{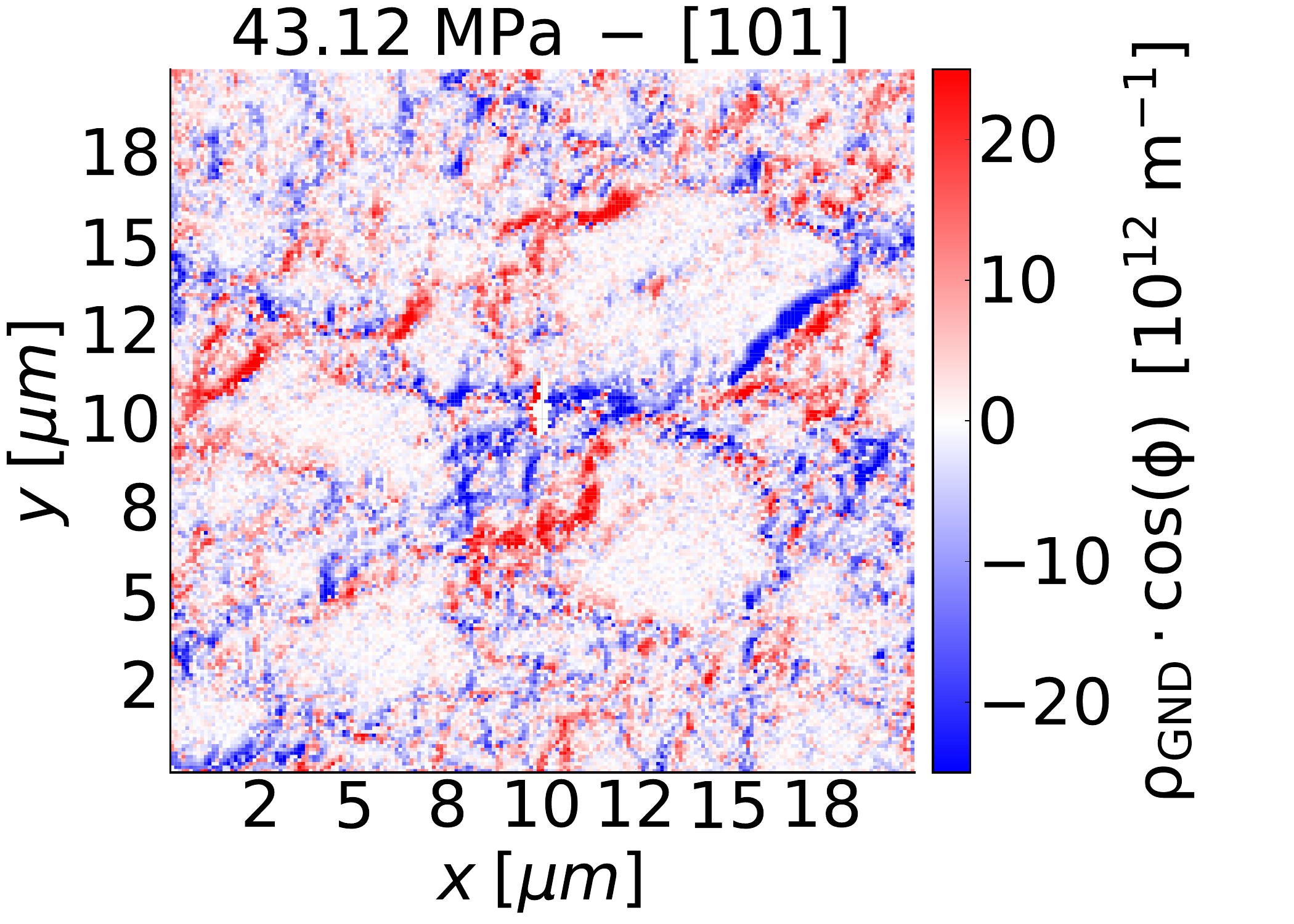}  \\
  \includegraphics[width=0.4\textwidth]{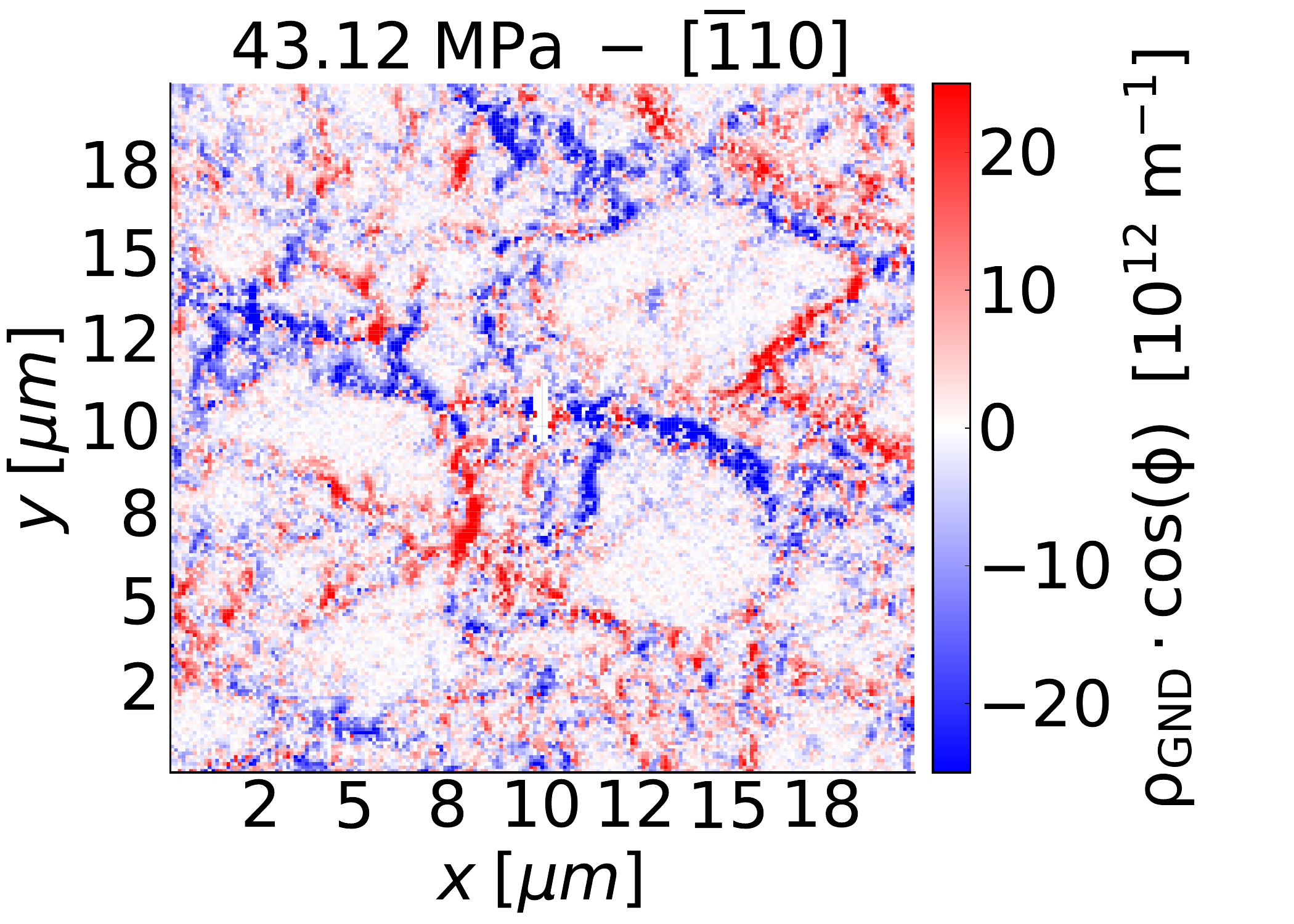}
  \includegraphics[width=0.4\textwidth]{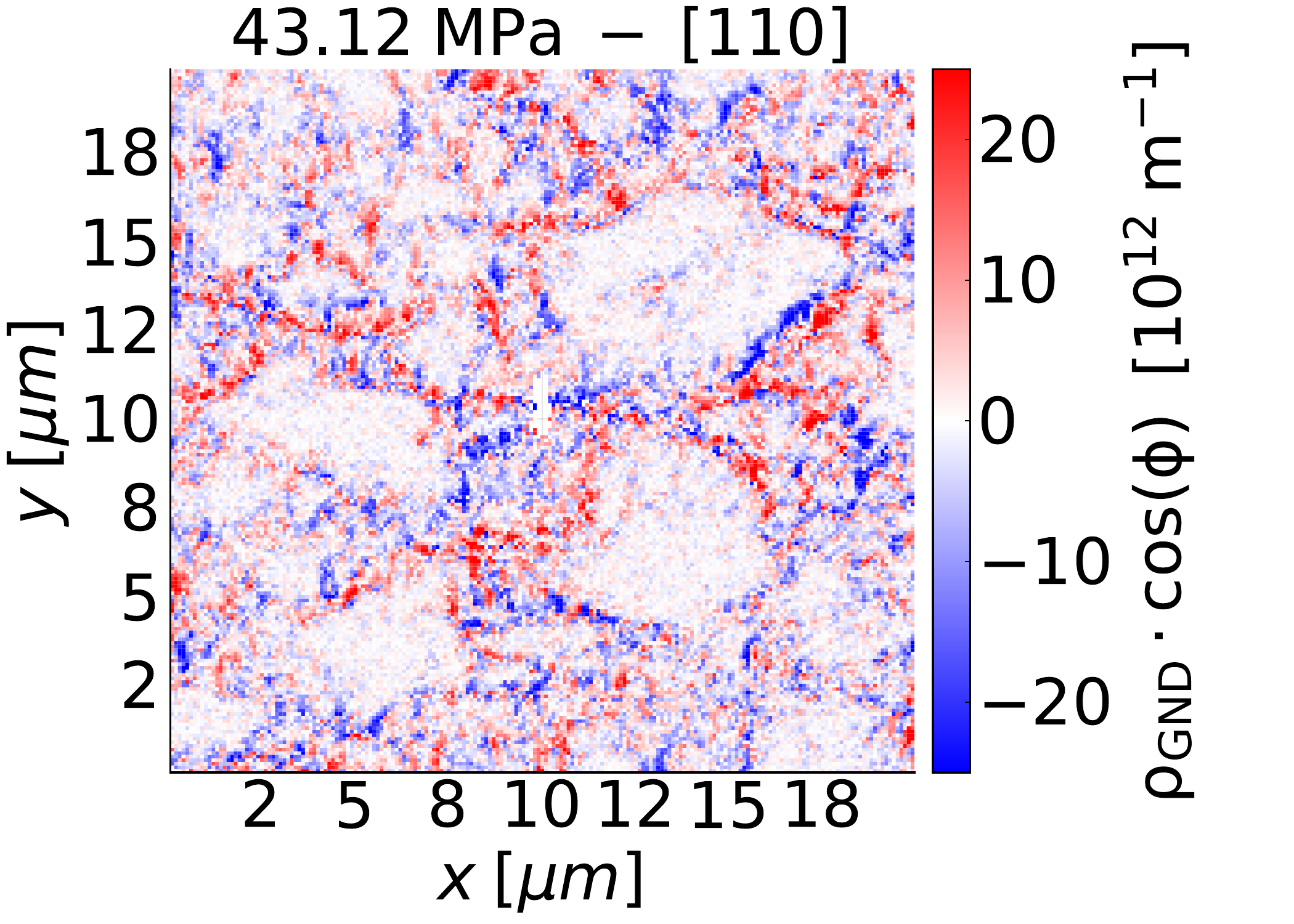} \\
   \includegraphics[width=0.4\textwidth]{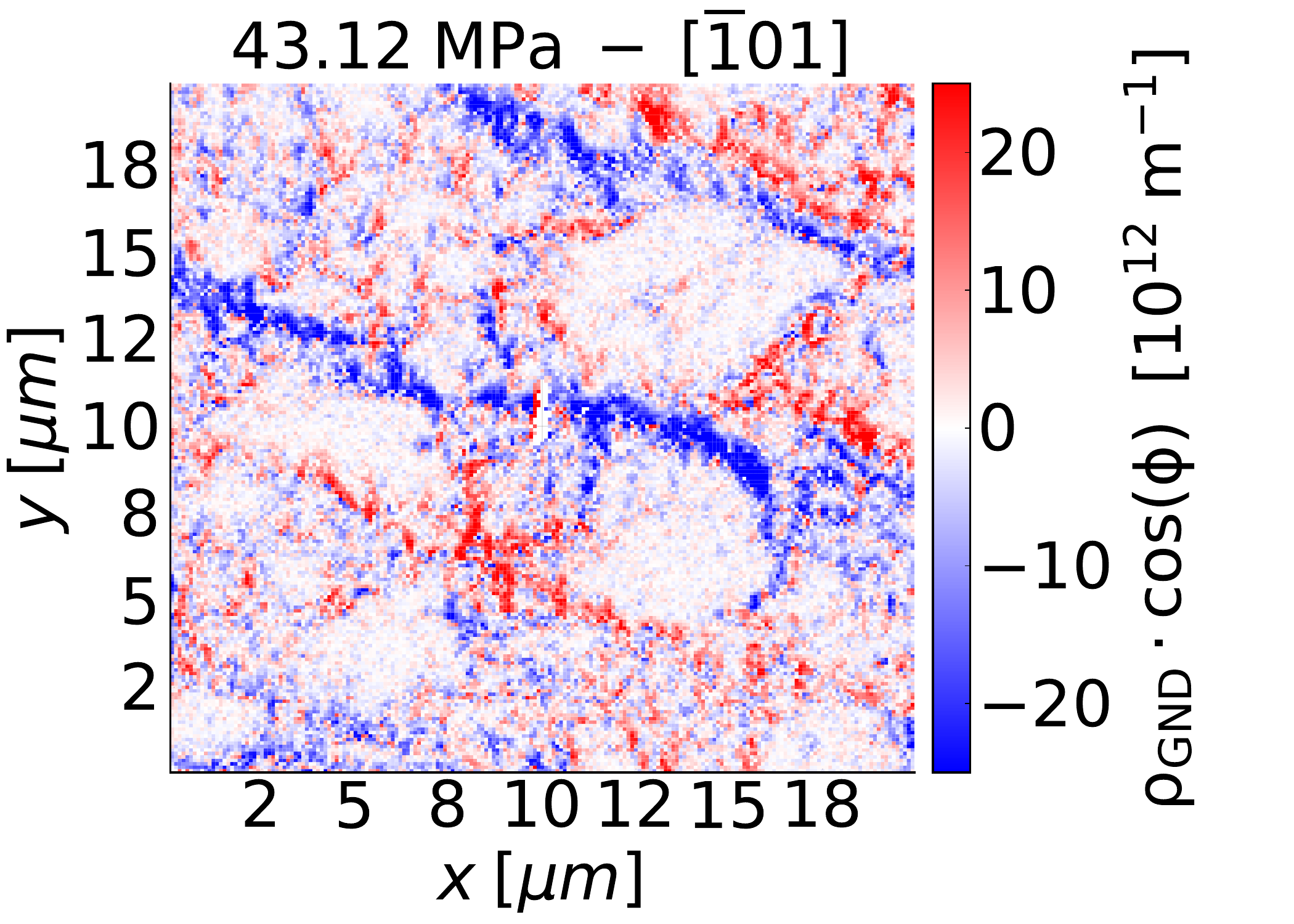}
  \includegraphics[width=0.4\textwidth]{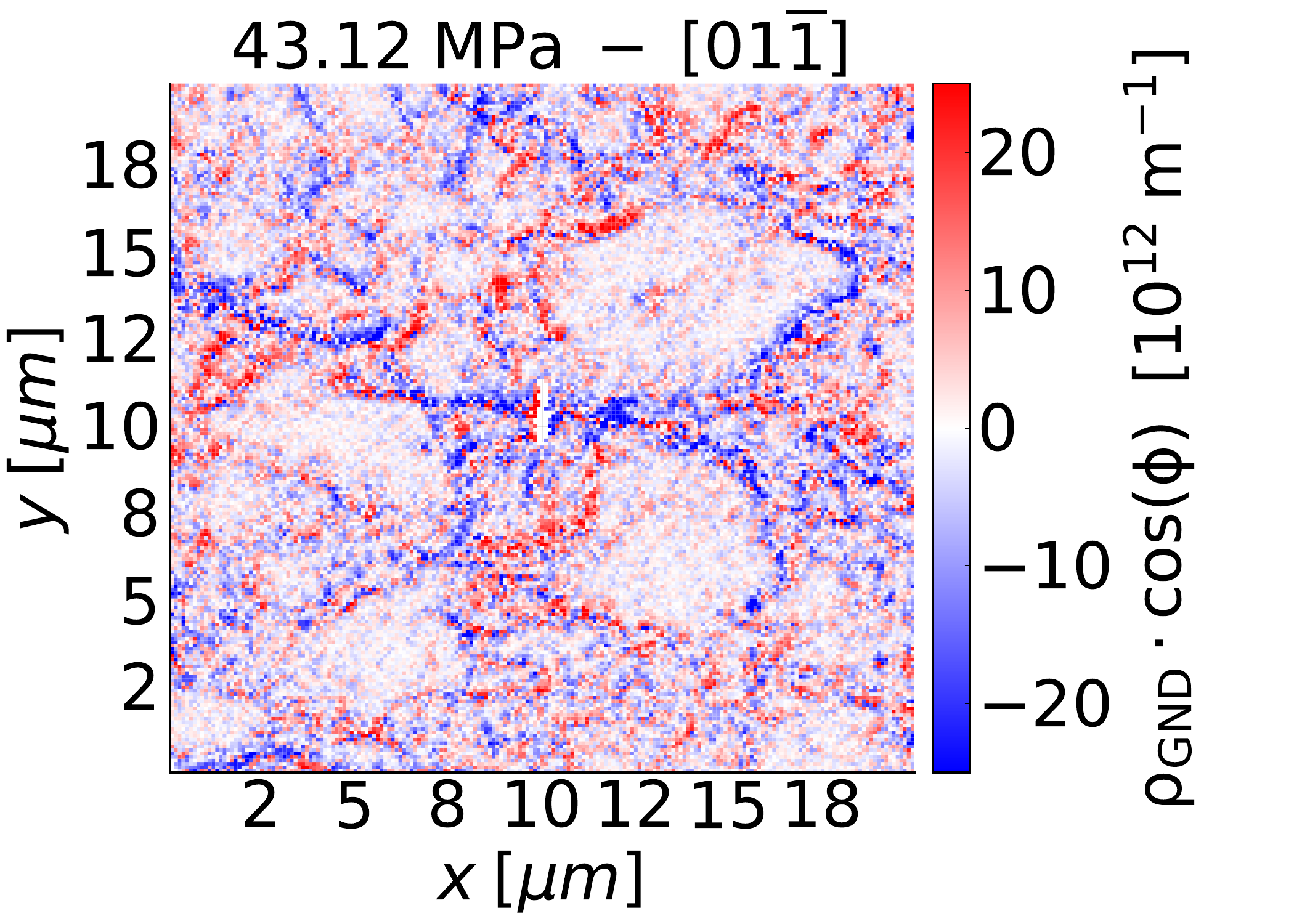}
	\caption{The $\rho_\mathrm{GND}\cdot a_i, \ \ i=1..6$ maps obtained on the sample compressed up to 43.12 MPa.}
  \label{BurgEval}
\end{figure}
In Sec.~\ref{sec:bv_analysis}, a method was outlined to analyze Burgers vectors based on the projection of 
vector $\vec{B}$ to the different possible Burgers vectors \cite{zoller2020microstructure}. For the sample deformed up to $43.12$ MPa the different $\rho_\mathrm{GND}\cdot a_i$ maps obtained are plotted in Fig.~\ref{BurgEval}. Similar behavior is found for the other samples. As it is seen some of the walls have a positive (red) or negative (blue) net Burgers vector while the other ones are more dipole like with a positive net Burgers vector on one side and a negative one on the other side. 

This picture somewhat refines the composite model proposed by Mughrabi et al.~\cite{mughrabi2002long,ungar1984x}. In the original form of the model the sources of the long-range stress are the two dislocation walls allocated on the two sides of a cell wall. The dislocation walls are formed by the reaction of dislocations in two slip systems resulting a Burgers vector parallel to the surface of the cell wall, i.e.,\ the GND structure imagined is dipole like. However, an elongated region with finite length $d$ having a net Burgers vector can also generate long-range internal stress within the connected $d\times d$ sized area. So, the source of the long-range stress are not necessary dipole-like walls as assumed earlier.

\section{Summary and conclusions}

Copper single crystals oriented for (100) ideal multiple slip  were compressed uniaxially up to different stress levels. The dislocation microstructure developing in the samples were studied by X-ray line profile analysis and HR-EBSD.

The main conclusions are:
\begin{itemize}
    \item It is shown that HR-EBSD offers a new efficient method to study   dislocation microstructure with much less sample preparation effort than TEM conventionally applied;
    \item According to X-ray line profile investigations on compressed Cu single oriented for ideal multiple slip the relative dislocation density fluctuation exhibits a sharp maximum at stage II to III transition stress level; 
    \item The dislocation cell structure is well described by a hole fractal with fractal dimension decreasing monotonically with the relative dislocation density fluctuation;
    \item The presence of the long-range internal stress developing in the cell interiors is directly seen by HR-EBSD measurements. Moreover, the stress maps can be directly measured. 
    \item Some of the walls have a positive or negative net Burgers vector while the other ones are more dipole-like with positive net Burgers vector on one side and negative one other other side. 
\end{itemize}
The results obtained can be directly compared to the prediction of the theoretical models, so they can help to inspire and validate them. 
Finally, it is important to emphasize that there are still a lot of issues that should be addressed. One very important question is related to size effect, especially the formation of cells in micropillars, that may lead to new interesting results. Moreover, it would be important to study in-situ the cell formation process.

\section*{Acknowledgments}

The  authors are grateful to Prof.~Géza Györgyi for the fruitful discussions on fractal analysis. 
This work has been supported by the National Research, Development and Innovation Office of Hungary
(PDI and IG, project No.~NKFIH-FK-138975). JLL acknowledges the support by VEKOP-2.3.3-15-2016-00002 of the European Structural and Investment Funds.


\begin{thebibliography}{10}
\expandafter\ifx\csname url\endcsname\relax
  \def\url#1{\texttt{#1}}\fi
\expandafter\ifx\csname urlprefix\endcsname\relax\def\urlprefix{URL }\fi
\expandafter\ifx\csname href\endcsname\relax
  \def\href#1#2{#2} \def\path#1{#1}\fi

\bibitem{hansen1986low}
N.~Hansen, D.~Kuhlmann-Wilsdorf, Low energy dislocation structures due to
  unidirectional deformation at low temperatures, Materials Science and
  Engineering 81 (1986) 141--161.

\bibitem{holt1970dislocation}
D.~L. Holt, Dislocation cell formation in metals, Journal of applied physics
  41~(8) (1970) 3197--3201.

\bibitem{rickman1997modelling}
J.~Rickman, J.~Vinals, Modelling of dislocation structures in materials,
  Philosophical Magazine A 75~(5) (1997) 1251--1262.

\bibitem{walgraef1985dislocation}
D.~Walgraef, E.~C. Aifantis, Dislocation patterning in fatigued metals as a
  result of dynamical instabilities, Journal of Applied Physics 58~(2) (1985)
  688--691.

\bibitem{pontes2006dislocation}
J.~Pont{\`e}s, D.~Walgraef, E.~Aifantis, On dislocation patterning: multiple
  slip effects in the rate equation approach, International Journal of
  Plasticity 22~(8) (2006) 1486--1505.

\bibitem{groma2016dislocation}
I.~Groma, M.~Zaiser, P.~D. Isp{\'a}novity, Dislocation patterning in a
  two-dimensional continuum theory of dislocations, Physical Review B 93~(21)
  (2016) 214110.

\bibitem{wu2018instability}
R.~Wu, D.~T{\"u}zes, P.~D. Isp{\'a}novity, I.~Groma, T.~Hochrainer, M.~Zaiser,
  Instability of dislocation fluxes in a single slip: Deterministic and
  stochastic models of dislocation patterning, Physical Review B 98~(5) (2018)
  054110.

\bibitem{wu2021cell}
R.~Wu, M.~Zaiser, Cell structure formation in a two-dimensional density-based
  dislocation dynamics model, Materials Theory 5~(1) (2021) 1--22.

\bibitem{hahner1998fractal}
P.~H{\"a}hner, K.~Bay, M.~Zaiser, Fractal dislocation patterning during plastic
  deformation, Physical Review Letters 81~(12) (1998) 2470.

\bibitem{zaiser1999flow}
M.~Zaiser, P.~H{\"a}hner, The flow stress of fractal dislocation arrangements,
  Materials Science and Engineering: A 270~(2) (1999) 299--307.

\bibitem{workhardening}
A.~Rollett, F.~Kocks, A review of the stages of work hardening, Solid State
  Phenomena 35 (1994) 1--18.

\bibitem{PhysRevB.57.7535}
I.~Groma, X-ray line broadening due to an inhomogeneous dislocation
  distribution, Phys. Rev. B 57 (1998) 7535--7542.

\bibitem{groma1988asymmetric}
I.~Groma, T.~Ung{\'a}r, M.~Wilkens, Asymmetric x-ray line broadening of
  plastically deformed crystals. i. theory, Journal of applied crystallography
  21~(1) (1988) 47--54.

\bibitem{Ungar:gk0172}
T.~Ung{\'{a}}r, I.~Groma, M.~Wilkens, {Asymmetric X-ray line broadening of
  plastically deformed crystals. II. Evaluation procedure and application to
  [001]-Cu crystals}, Journal of Applied Crystallography 22~(1) (1989) 26--34.

\bibitem{Wilkinson.2012}
T.~B. Britton, A.~J. Wilkinson, High resolution electron backscatter
  diffraction measurements of elastic strain variations in the presence of
  larger lattice rotations, Ultramicroscopy 114 (2012) 82--95.

\bibitem{kroner1981continuum}
E.~Kr{\"o}ner, et~al., Continuum theory of defects, Physics of defects 35
  (1981) 217--315.

\bibitem{Kalacska.2017}
S.~Kal{\'a}cska, I.~Groma, A.~Borb{\'e}ly, P.~D. Isp{\'a}novity, Comparison of
  the dislocation density obtained by {HR-EBSD} and {X}-ray profile analysis,
  Applied Physics Letters 110~(9) (2017) 091912.

\bibitem{Kalacska.2020a}
S.~Kal{\'a}cska, Z.~Dankh{\'a}zi, G.~Zilahi, X.~Maeder, J.~Michler, P.~D.
  Isp{\'a}novity, I.~Groma, Investigation of geometrically necessary
  dislocation structures in compressed {C}u micropillars by 3-dimensional
  {HR-EBSD}, Materials Science and Engineering: A 770 (2020) 138499.

\bibitem{groma1998probability}
I.~Groma, B.~Bak{\'o}, Probability distribution of internal stresses in
  parallel straight dislocation systems, Physical Review B 58~(6) (1998) 2969.

\bibitem{csikor2004probability}
F.~F. Csikor, I.~Groma, Probability distribution of internal stress in relaxed
  dislocation systems, Physical Review B 70~(6) (2004) 064106.

\bibitem{wallis2021dislocation}
D.~Wallis, L.~N. Hansen, A.~J. Wilkinson, R.~A. Lebensohn, Dislocation
  interactions in olivine control postseismic creep of the upper mantle, Nature
  Communications 12~(1) (2021) 1--12.

\bibitem{mayer2007tem}
J.~Mayer, L.~A. Giannuzzi, T.~Kamino, J.~Michael, {TEM} sample preparation and
  {FIB}-induced damage, MRS bulletin 32~(5) (2007) 400--407.

\bibitem{giannuzzi2004introduction}
L.~A. Giannuzzi, et~al., Introduction to focused ion beams: instrumentation,
  theory, techniques and practice, Springer Science \& Business Media, 2004.

\bibitem{yao2007focused}
N.~Yao, Focused ion beam systems: basics and applications, Cambridge University
  Press, 2007.

\bibitem{salat2017multifractal}
H.~Salat, R.~Murcio, E.~Arcaute, Multifractal methodology, Physica A:
  Statistical Mechanics and its Applications 473 (2017) 467--487.

\bibitem{grassberger2004measuring}
P.~Grassberger, I.~Procaccia, Measuring the strangeness of strange attractors,
  in: The theory of chaotic attractors, Springer, 2004, pp. 170--189.

\bibitem{otsu1979threshold}
N.~Otsu, A threshold selection method from gray-level histograms, IEEE
  transactions on systems, man, and cybernetics 9~(1) (1979) 62--66.

\bibitem{lee1990comparative}
S.~U. Lee, S.~Y. Chung, R.~H. Park, A comparative performance study of several
  global thresholding techniques for segmentation, Computer Vision, Graphics,
  and Image Processing 52~(2) (1990) 171--190.

\bibitem{fisher1936use}
R.~A. Fisher, The use of multiple measurements in taxonomic problems, Annals of
  Eugenics 7~(2) (1936) 179--188.

\bibitem{macqueen1967some}
J.~MacQueen, et~al., Some methods for classification and analysis of
  multivariate observations, in: Proceedings of the fifth Berkeley symposium on
  mathematical statistics and probability, Vol. 1(14), Oakland, CA, USA, 1967,
  pp. 281--297.

\bibitem{kriegel2017black}
H.-P. Kriegel, E.~Schubert, A.~Zimek, The (black) art of runtime evaluation:
  Are we comparing algorithms or implementations?, Knowledge and Information
  Systems 52~(2) (2017) 341--378.

\bibitem{zoller2020microstructure}
K.~Zoller, S.~Kal{\'a}cska, P.~D. Isp{\'a}novity, K.~Schulz, Microstructure
  evolution of compressed micropillars investigated by in situ {HR-EBSD}
  analysis and dislocation density simulations, Comptes Rendus Physique, Spec.
  Iss. Plasticity \& Solid State Physics (2021) 1--27.

\bibitem{szekely2001statistic}
F.~Szekely, I.~Groma, J.~Lendvai, Statistic properties of dislocation
  structures investigated by x-ray diffraction, Materials Science and
  Engineering: A 309 (2001) 352--355.

\bibitem{mughrabi2002long}
H.~Mughrabi, T.~Ung{\'a}r, Long-range internal stresses in deformed
  single-phase materials: the composite model and its consequences, in:
  Dislocations in solids, Vol.~11, Elsevier, 2002, pp. 343--411.

\bibitem{ungar1984x}
T.~Ungar, H.~Mughrabi, D.~R{\"o}nnpagel, M.~Wilkens, X-ray line-broadening
  study of the dislocation cell structure in deformed [001]-orientated copper
  single crystals, Acta Metallurgica 32~(3) (1984) 333--342.

\bibitem{oudriss2016length}
A.~Oudriss, X.~Feaugas, Length scales and scaling laws for dislocation cells
  developed during monotonic deformation of (001) nickel single crystal,
  International Journal of Plasticity 78 (2016) 187--202.

\end{thebibliography}





\end{document}